\newcommand{\beq}{\begin{equation}}
\newcommand{\eeq}{\end{equation}}
\newcommand{\bea}{\begin{eqnarray}}
\newcommand{\eea}{\end{eqnarray}}

\newcommand{\gsim}{\lower.7ex\hbox{$\;\stackrel{\textstyle>}{\sim}\;$}}
\newcommand{\lsim}{\lower.7ex\hbox{$\;\stackrel{\textstyle<}{\sim}\;$}}

\documentclass[a4paper,11pt]{article}
\pdfoutput=1

\usepackage{jcappub}
\usepackage[utf8]{inputenc} 
\usepackage{amsfonts,amssymb,mathrsfs,amsmath,esint,bm}
\usepackage{pdflscape} 
\usepackage[capitalise]{cleveref}
\allowdisplaybreaks

\graphicspath{{./Figures/}}
\usepackage{latexsym}
\usepackage{graphicx}
\usepackage{subfigure}
\usepackage[dvipsnames]{xcolor}
\usepackage{booktabs}
\usepackage{physics}
\usepackage{tikz}
\usetikzlibrary{decorations.pathmorphing}\usetikzlibrary{positioning,calc}
\usetikzlibrary{math}
\usepackage{cleveref}
\usepackage{comment}
\usepackage[normalem]{ulem}

\hypersetup{pdfstartview=FitV,colorlinks=true,linkcolor=blue,citecolor=red,filecolor=black,urlcolor=blue}

\def\stacksymbols #1#2#3#4{\def\theguybelow{#2}
    \def\vp{\lower#3pt}
    \def\sp{\baselineskip0pt\lineskip#4pt}
    \mathrel{\mathpalette\intermediary#1}}

\def\intermediary#1#2{\vp\vbox{\sp
     \everycr={}\tabskip0pt
     \halign{$\mathsurround0pt#1\hfil##\hfil$\crcr#2\crcr
              \theguybelow\crcr}}}

\newcommand{\ep}{\epsilon}

\newcommand{\al}{\alpha}

\newcommand{\yob}{\bar{y}_1}
\newcommand{\ytb}{\bar{y}_2}

\title{Radiative Corrections in Supergravity Models of Inflation}

\author[a,b]{John Ellis,}
\affiliation[a]{Theoretical Particle Physics and Cosmology Group, Department of
  Physics, King's~College~London, London WC2R 2LS, United Kingdom}
\affiliation[b]{Theoretical Physics Department, CERN, CH-1211 Geneva 23,
  Switzerland}
\emailAdd{john.ellis@cern.ch}

\author[c]{Tony Gherghetta,}
\affiliation[c]{School of Physics and Astronomy, University of Minnesota, Minneapolis, MN 55455, USA}
\emailAdd{tgher@umn.edu}

\author[d]{Kunio Kaneta,}
\affiliation[d]{Faculty of Education, Niigata University, Niigata 950-2181, Japan}
\emailAdd{kaneta@ed.niigata-u.ac.jp}

\author[e]{Wenqi Ke,}
\affiliation[e]{William I.~Fine Theoretical Physics Institute, School of Physics and Astronomy, University of Minnesota, Minneapolis, MN 55455, USA}
\emailAdd{wke@umn.edu} 

\author[e]{~and~Keith A. Olive,}
\emailAdd{olive@umn.edu}

\abstract{Supergravity provides the natural supersymmetric framework for early universe cosmology. A broad class of inflationary models in no-scale supergravity yields tree-level predictions for cosmic microwave background (CMB) observables that closely resemble those of the Starobinsky $R + R^2$ model. Using results from global supersymmetry and supergravity, we analyze radiative corrections in models with canonical and non-canonical kinetic terms, focusing particularly on Starobinsky-like no-scale supergravity models. 
We derive conditions on the superpotential that keep the gravitino mass finite during inflation and ensure that loop-induced corrections to the K\"ahler potential remain either finite or subdominant relative to the tree-level potential.  
We show that in some models, most notably the original no-scale supergravity model with a Wess-Zumino superpotential, radiative corrections grow at large inflaton field values and can dominate the inflationary dynamics, rendering unreliable the model predictions for CMB data. However, we identify a class of no-scale Starobinsky-like models, including the Cecotti model, in which radiative corrections remain very small for inflaton field values $\lesssim 8$ (in Planck units), preserving the agreement of the tree-level predictions with Planck CMB data.
}

\begin{document}
\begin{flushright}
UMN--TH--4521/26, FTPI--MINN--26/05,   \\
KCL--PH--TH/2026-05, CERN--TH--2026-022 \\
March  2026
\end{flushright}
\maketitle

\section{Introduction}

The advent of more precise measurements \cite{Planck} of the cosmic microwave background (CMB) motivates aiming for greater precision in evaluating the predictions of inflation models. A model that has remained remarkably consistent with CMB observations is the original Starobinsky model based on an $R + R^2$ extension of the minimal Einstein-Hilbert action \cite{Staro}. This success has generated considerable interest in field-theoretical models that can reproduce ~\cite{eno6,Kallosh:2013lkr,Ellis:2013nxa,Ellis:2018zya} CMB predictions derived from the Starobinsky potential
\beq
V = \frac34 M^2 \left(1-e^{-\sqrt{\frac23}\phi}\right)^2
\label{staropot} \, , 
\eeq
frequently at the expense of fine-tuning model parameters~\cite{Antoniadis:2025pfa}. Recent hints of possible observational deviations \cite{ACT:2025fju} from the predictions of the Starobinsky model have stimulated studies of possible deformations of these field-theoretical avatars \cite{Kallosh:2025rni,Dioguardi:2025vci,Gialamas:2025kef,Antoniadis:2025pfa,Drees:2025ngb,Zharov:2025evb,Haque:2025uri,Gialamas:2025ofz,Haque:2025uga,Wolf:2025ecy,Ahmed:2025rrg,Han:2025cwk,Pallis:2025gii,German:2025ide,deform,Aoki:2025ywt,Ellis:2025bzi,Ellis:2025zrf,Alexandre:2025ixz,Iacconi:2025odq,Kallosh:2025sji}. Among these have been calculations of radiative corrections to the tree-level predictions of field-theoretical models \cite{Gialamas:2025kef,Wolf:2025ecy,Ahmed:2025rrg,Han:2025cwk,Ellis:2025bzi,Alexandre:2025ixz}. In this paper we extend these investigations to supersymmetric models, focusing in particular on models formulated in the framework of supergravity.

Supersymmetry has two key features that are desirable for inflationary model building \cite{ENOT,Nakayama:2011ri}: it can stabilize the inflaton mass against large radiative corrections, and it can preserve the fine-tuning of parameters that are often required in many field-theoretical avatars of the Starobinsky model. When supersymmetry is combined with gravity, supergravity provides the natural framework for analyzing inflation and other issues in cosmology \cite{Nanopoulos:1982bv,hrr}. In particular, no-scale supergravity \cite{no-scale1,EKN,ELNT,no-scale2} is especially appealing because it ensures that the effective potential is positive semi-definite at tree level. Furthermore, no-scale structures arise naturally as the effective four-dimensional low-energy theory in string theory \cite{Witten}.

In this paper we first study radiative corrections in a toy globally-supersymmetric theory with a canonical Hermitian quadratic K\"ahler function $K$, corresponding to a flat Kähler metric with minimal quadratic derivative terms, and a holomorphic superpotential $W$.  Examination of the one-loop correction to the Kähler potential reveals logarithmically-divergent terms that correspond to the expected multiplicative renormalization of Yukawa couplings. We also show that the same one-loop correction to the K\"ahler potential reproduces the familiar one-loop Coleman-Weinberg (CW) contribution to the effective scalar potential.

Following this warm-up exercise, we turn to supergravity, taking as a starting point a tree-level theory with suitably tuned higher-derivative and/or stabilizing terms to realize new inflation, or a Starobinsky plateau. By computing the one-loop corrections, in particular to the K\"ahler potential, we can then test the radiative stability of the resulting inflationary predictions. Importantly, we compute the radiative corrections in Minkowski space, since the spacetime curvature (de Sitter) effects are negligible, as is justified in Section~\ref{sec:sugra}. We begin with two supergravity inflationary models with canonical quadratic K\"ahler functions, one with a single chiral field $\Phi$ with $K= \Phi \bar{\Phi}  $ and superpotential $W = M(\Phi - 1)^2$ \cite{hrr} and the other with two chiral fields $(\Phi, S)$ and $K=S\bar{S}-\frac{1}{2}\left(\Phi-\bar{\Phi}\right)^2$ and a superpotential of the form $S \Phi$  \cite{Kawasaki:2000yn}. The former is an example of new inflation \cite{Linde:1981mu} and the latter of chaotic inflation \cite{Linde:1983gd}. We find that the one-loop corrections do not bring the predictions of these models into agreement with the CMB data. However, the construction used in~\cite{Kawasaki:2000yn} can be generalized to generate any potential that is a perfect square, including the Starobinsky model~\cite{Kallosh:2010ug,Kallosh:2010xz,Kallosh:2011qk}. In this case the radiative corrections are small, and preserve agreement with the CMB data.

Next, we consider no-scale supergravity models that reproduce the CMB predictions of the Starobinsky model at tree level, and are formulated as SU(2,1)/SU(2)$\times$U(1) coset field theories \cite{eno6,Ellis:2013nxa,Cecotti,Ellis:2018zya,building}. The coset may be parametrized in terms of a generalized modulus field $T$ and a matter field $\chi$, with Kähler function $K = -3 \ln(T + \bar{T} - |\chi|^2/3)$. Inflationary models may be constructed by stabilizing dynamically one of these fields and identifying the other as the inflaton. 
For example, the original no-scale supergravity model \cite{eno6} assumed $T$ to have a fixed non-zero value, postulated a minimal Wess-Zumino superpotential for $\chi$, and identified $\chi$ as the inflaton field. Another interesting Starobinsky-like model, proposed in~\cite{Cecotti} in a different context, can be written with the same Kähler function but a different superpotential $W = \sqrt{3} M \chi (T - 1/2)$, which we refer to as the Cecotti model. In this case, $\chi$ is assumed to be at zero, while $T$ can be identified as the inflaton field. 

A general classification of Starobinsky-like inflationary models in the framework of SU(2,1)/SU(2)$\times$U(1) no-scale supergravity and their relationships has been given in~\cite{Ellis:2018zya}. For our purposes, it is convenient to use a symmetric parametrization of the SU(2,1)/SU(2)$\times$U(1) manifold in terms of a pair of chiral fields $y_{1,2}$ with $K = -3 \ln\left(1 - (|y_1|^2 + |y_2|^2)/3 \right)$.

We find that the radiative corrections to no-scale Starobinsky-like inflationary potentials are generally small for inflaton field values $\lesssim 6$.~\footnote{Throughout, we work in Planck units setting $M_P^2 \equiv 1/8\pi G_{\rm N}=1$. } However, there are two circumstances under which the radiative corrections can become important at larger inflaton field values, leading to a breakdown of the perturbative expansion, making the CMB predictions unreliable.
One possibility is that supersymmetry breaking grows without bound for large values of the inflaton field, which would enhance the loop contributions to the effective potential. This problem is avoided in the Cecotti model and in a broader class of generalizations with superpotential $W = y_2~u(y_1, y_2)$, where $u(y_1,y_2)$ is an arbitrary function. A second possibility is that the radiative correction to the Kähler potential develops a higher-order singularity than the tree-level potential, again leading to the breakdown of the perturbation expansion at large field values. This second problem is likewise avoided for a restricted form of $u(y_1, y_2)$, as realized in a suitable generalization of the Cecotti model.

The outline of the paper is as follows. 
In Section \ref{sec1}, we first consider the one-loop corrections in global supersymmetry. By computing the corrections to the kinetic terms (i.e., to a minimal K\"ahler potential), we extract the $\beta$-functions for all superpotential couplings and obtain the associated one-loop correction to the scalar potential. In Section \ref{sec:sugra}, we then turn to supergravity, using the general results of \cite{Gaillard:1993es,Gaillard:1996hs} for the one-loop corrections to the K\"ahler potential in supergravity models. In Section \ref{new}, we apply these corrections to two illustrative models based on minimal supergravity: a new inflation model \cite{hrr} and one employing a shift symmetry in the K\"ahler potential \cite{Kawasaki:2000yn}. In Section \ref{nssugra}, we first review the construction of no-scale Starobinsky avatars as models of inflation and then derive the conditions on the superpotential that ensure that the gravitino mass remains finite during inflation. We also emphasize that the K\"ahler potential is generically singular as the canonical inflaton is taken to large field values. We derive conditions on the superpotential that avoid these issues, so as to guarantee that loop corrections to the K\"ahler potential remain either finite or subdominant. Finally, in Section \ref{oneloopav} we examine two well-studied no-scale supergravity models of inflation -- the no-scale Wess-Zumino model~\cite{eno6} and the Cecotti model~\cite{Cecotti} -- in light of these conditions. Our summary and conclusions are presented in Section \ref{sec:conx}, while in Appendix~\ref{app:k1} we explicitly write out the complete expressions for the one-loop corrections to the K\"ahler potential for various models.

\section{One-loop Effective Potential in Global and Local Supersymmetry}

In this Section, we first review the one-loop effective potential in global supersymmetry, highlighting the relation between the one-loop K\"ahler correction and the Coleman-Weinberg potential, and showing explicitly that both approaches satisfy the Callan-Symanzik equation at one loop. We then consider local supersymmetry (supergravity), where the one-loop structure is more involved because of gravitational and curvature effects. In particular, we focus on the logarithmically-divergent terms, which will be used to calculate the renormalization-improved one-loop potential in supergravity.

\subsection{Global Supersymmetry}
\label{sec1}

The one-loop correction to the Kähler potential is derived by expanding the action to quadratic order around some background superfield, and then integrating out the quantum fluctuations  
\cite{Grisaru:1996ve,Buchbinder:1998twe}. If the tree-level Kähler potential is \textit{canonical}, i.e., $K^{(0)}= \sum _ i |\Phi_i|^2$,  then in Landau gauge the one-loop correction is
\begin{equation}
    K^{(1)}=-\frac{1}{32\pi ^2} \text{Tr}\left[ \bar{M}M \left(\log\frac{\bar{M}M}
    {\mu^2}-1\right)\right] \, ,
    \label{minoneloop}
\end{equation}
where $\bar M\equiv M^\dagger$, $\mu$ is a renormalization scale and, for simplicity, we assume throughout this work that no vector multiplets are present. The ‘‘superfield-dependent'' mass $M$ is given by
\begin{equation}
    M_{ij}\equiv \partial_{\Phi_i}\partial_{\Phi_j}W \, .
\end{equation}
As a toy example, in the massless Wess-Zumino model $W=y \Phi^3/6$, where $y$ is a dimensionless Yukawa coupling, the one-loop Kähler potential is
\begin{equation}
     K^{(1)}=-\frac{1}{32\pi ^2} |y\Phi|^2\left(\log\frac{|y\Phi|^2}{\mu^2}-1\right) \, .
     \label{one-loopKsusy}
\end{equation}
This yields the following correction to the Kähler metric
\begin{equation}
    K^{(1)}_{\bar{\Phi}\Phi}=-\frac{1}{32\pi ^2}|y|^2\left(\log\frac{|y\Phi|^2}{\mu^2}+1\right) \, .
    \label{k1global}
\end{equation}
The $\mu $-dependent part of the scalar kinetic term extracted from (\ref{one-loopKsusy}) yields the wavefunction renormalization factor 
\begin{equation}
    \delta Z(\mu) =\frac{1}{32\pi^2}|y|^2 \log \mu^2 \, , 
\end{equation}
from which we obtain the well-known scalar anomalous dimension
\begin{equation}
    \gamma_\phi=\frac{|y|^2}{32\pi^2} \, .
    \label{eqgamma}
\end{equation}
Returning to the superpotential, the non-renormalization theorem implies that the running of $y$ arises entirely from the wavefunction renormalization, i.e., $y_\text{bare}=y_\text{phys}Z^{-3/2}$, which then determines the $\beta$-function for $y$
\begin{equation}
    \beta_y=3\gamma_\phi y\,.
\end{equation}

We can verify that the Callan-Symanzik (CS) equation is satisfied at one loop. Writing the complex scalar component of the chiral superfield $\Phi$ as $\phi=(\phi_R+i \phi_I)/\sqrt{2}$, we assume a background field value along the real direction, $\phi_I=0$ so that $\langle \phi\rangle=\phi_0\equiv\phi_R/\sqrt{2}$. Assuming for simplicity that $y$ is real, the two real scalars have mass eigenvalues 
\begin{equation}
    m_\pm^2=y^2\phi_0^2 \left(1\pm \frac{1}{2}\right) \, ,
\end{equation}
and the fermion partner has mass $m_\psi = y \phi_0$.\footnote{Note that, for $\phi_0=0$, all fields are massless and supersymmetry is unbroken.} The one-loop CW correction to the potential is 
\begin{equation}
\begin{aligned}
    V_1(\phi_0)&=\frac{1}{64\pi^2} {\rm Str}\Bigg\{ \mathcal{M}_i ^4 \left(\log \frac{\mathcal{M}_i ^2}{\mu^2}-\frac{3}{2}\right) \Bigg\}\,,   \\&=\frac{1}{64\pi^2}\left[m_+^4\left(\log\frac{m_+^2}{\mu^2}-\frac{3}{2}\right)+m_-^4\left(\log\frac{m_-^2}{\mu^2}-\frac{3}{2}\right)-2m_\psi^4\left(\log\frac{m_\psi^2}{\mu^2}-\frac{3}{2}\right) \right],
\end{aligned}
\label{v1phi0}
\end{equation}
where ${\rm Str}\{...\}$ denotes the supertrace, i.e., a sum over all degrees of freedom with factors of $+1 (-1)$ for each bosonic (fermionic) degree of freedom. Differentiating with respect to $\log\mu$ gives
\begin{equation}
  \left.    \frac{\partial V_1}{\partial \log\mu }\right|_{\phi=\phi_0}=-\frac{1}{64\pi^2} y^4 \phi_0^4 \, .
  \label{dv1global}
\end{equation}
For comparison, acting on the tree-level potential $V_0$ yields
\begin{equation}
  \left.  \left(\beta_y\frac{\partial}{\partial y}  -\gamma _\phi \phi \frac{\partial}{\partial \phi}\right)V_0 \right|_{\phi=\phi_0}= \left.  \left(3\gamma _\phi y\frac{\partial}{\partial y}  -\gamma _\phi \phi \frac{\partial}{\partial \phi}\right)V_0 \right|_{\phi=\phi_0}=\frac{1}{2}y^2\phi_0^4\gamma_\phi \, .
\end{equation}
Using Eq.~\eqref{eqgamma} and Eq.~\eqref{dv1global} the CS equation
\begin{equation}
    \left( \frac{\partial  }{\partial \log\mu }+\beta_y\frac{\partial}{\partial y}  -\gamma _\phi \phi \frac{\partial}{\partial \phi}\right)(V_0+V_1)=0\,,
\end{equation}
is satisfied at one loop.  

We emphasize that, in the above calculation, the one-loop corrected potential is the sum of $V_0$ -- the tree-level potential obtained from the \textit{tree-level} Kähler ($K^{(0)}$) and the superpotential ($W$), and $V_1$ -- the one-loop CW correction. The one-loop CW contribution \eqref{v1phi0} is required for the CS equation to hold at this order. 

Alternatively, one may obtain the one-loop corrected potential directly from the \textit{one-loop corrected} Kähler potential ($K=K^{(0)}+K^{(1)}$), together with the superpotential ($W$), using the standard globally-supersymmetric expression
\begin{equation}
    V_F= K^{\Phi \bar{\Phi}}W_\Phi \bar{W}_{\bar{\Phi}}\,.
\label{vfglobal}
\end{equation}
In this single-field example, 
\begin{equation}
    K^{\Phi \bar{\Phi}}=(1+K^{(1)}_{\Phi \bar{\Phi}})^{-1} \, ,
\end{equation}
and absorbing the constant term in \eqref{k1global} into a redefinition of the renormalization scale $\mu$, the resulting globally-supersymmetric potential (\ref{vfglobal}) becomes
\begin{equation}
    V_F(\phi_0)=\frac{1}{4}\frac{y^2\phi_0^4 }{1- \frac{y^2}{32\pi^2}  \log(y^2\phi_0^2   /\mu ^2)} \, .
\end{equation}
Expanding the denominator assuming that the logarithm is small, we recover the expected tree-level potential $V_0=y^2\phi_0^4/4$ and the one-loop correction
\begin{equation}
    V_1=\frac{  y^4\phi_0^4  }{128 \pi ^2 } \log(  y^2\phi_0^2   /\mu ^2) \, .
    \label{v1kal}
\end{equation}
It is straightforward to verify that Eq.~\eqref{dv1global} holds, and hence the CS equation is satisfied at one loop.

To summarize, we have two ways to calculate the one-loop corrected potential: (1) the tree-level potential plus the CW correction in Eq.~(\ref{v1phi0}), and (2) the globally-supersymmetric expression \eqref{vfglobal} evaluated with the one-loop corrected Kähler potential, which reproduces the same tree-level term and yields the one-loop correction in Eq.~(\ref{v1kal}). Both approaches satisfy the CS equation. Comparing the two expressions for $V_1$, we note that Eq.~\eqref{v1phi0} gives (dropping the constant $3/2$, for simplicity)
\begin{equation}
    V_1(\phi_0)=\frac{1}{64\pi^2}\left(   - \frac{1}{2} y^4\phi_0^4\log  \mu^2 +m_+^4\log m_+^2+m_-^4\log m_-^2-2m_\psi^4\log m_\psi^2
    \right) \, .
    \label{v1no32}
\end{equation}
The $\log\mu^2$-dependent term matches with Eq.~\eqref{v1kal} (which is why both methods satisfy the CS equation). However, the remaining mass-dependent logarithms in Eq.~(\ref{v1no32}) differ from Eq.~\eqref{v1kal} because $m_\psi,m_+$ and $ m_-$ are not equal. In general, when the mass splittings are small compared to the largest mass scale in the background, the effective potential obtained by the one-loop Kähler potential becomes a good approximation to the CW result \eqref{v1phi0}. However, when the supersymmetry-breaking effects are sizeable, additional contributions (e.g., from the effective auxiliary-field action \cite{Martin:2024qmi}) are required to reproduce the full one-loop scalar effective potential. 

\subsection{Supergravity}
\label{sec:sugra}

In supergravity, the one-loop structure is more complicated due to graviton loops and the coupling to spacetime curvature, so \eqref{v1phi0} is not directly applicable. In \cite{Gaillard:1993es}, the \textit{logarithmically} divergent  part of the one-loop effective action in supergravity was derived by evaluating the trace $\log$ expression. The result can be written as an effective Lagrangian $\mathcal{L}(K_R,g_R)$ expressed in terms of the renormalized Kähler potential $K_R$ and metric $g_R$, in addition to Planck-suppressed terms. 
In the globally-supersymmetric limit, $M_P\rightarrow\infty$, the effective potential reduces to the tree-level potential with the one-loop corrected Kähler potential, namely, the potential obtained via method (2) discussed in the previous section. Consequently, the $\log\mu^2$-dependent part agrees with the globally-supersymmetric result in this limit. We therefore focus on the $\log\mu^2$-dependent one-loop correction in supergravity, using Eq.~(3.6) from~\cite{Gaillard:1993es}.

As we will see later, when we study specific examples of inflationary models in supergravity, the  one-loop Kähler potential and the one-loop scalar potential can diverge at large field values. The starting point of the deformation with respect to the tree-level potential affects the CMB predictions, so this procedure should be made as scale-independent as possible. This can be done by introducing counterterms and enforcing the CS equation. Since supergravity is non-renormalizable, absorbing these divergences generally requires an infinite number of counterterms. We encode this set of counterterms in a function $V_{\rm CT}(\mu,\phi)$, which is taken to vanish at the initial scale $\mu_0$. Requiring $\mu$-independence  at one-loop order then implies
\begin{equation}
    \frac{dV_{\rm RGI}(\mu,\phi)}{d\log\mu }=\sum_{k} \beta_{\lambda_k}\frac{\partial V_0}{\partial \lambda_k}-\sum _{i,j}\gamma_{\phi_i\bar{\phi}_j} \phi_i\frac{\partial V_0}{\partial\bar{\phi}_j}+\frac{\partial V_1}{\partial \log\mu }+\frac{\partial V_{\rm CT}}{\partial \log \mu }=0\,,
\end{equation}
where $V_0$ is the tree-level potential and $V_1$ is the  one-loop potential given in \cite{Gaillard:1993es}. For each value of 
$\phi$ we obtain a ‘‘$\beta$-function'' for $V_{\rm CT}$, and integrating this function, we obtain the value of $V_{\rm CT}$ at $\phi$.  The total  potential is  then $V_{\rm RGI} =V_0+V_1+V_{\rm CT}$ with running couplings.

For general (non-minimal) K\"ahler potentials and local supersymmetry, the one-loop correction to the K\"ahler potential, ignoring the contribution from vector multiplets, can be expressed as  \cite{Gaillard:1993es,Gaillard:1996hs}
\begin{equation}
    K^{(1)}=\frac{\log\mu^2}{32\pi ^2}  e^{-K}\left[A_{ij}\bar{A}^{ij}-2A_{i}\bar{A}^{i}-4A \bar{A} \right] \, ,
    \label{Kgeneral}
\end{equation}
where $A, \bar A$ are defined by
\begin{equation}
    A\equiv e^K W\,, \quad \bar{A}\equiv e^K \bar{W}\,,\label{defa}
\end{equation}
with
\begin{equation}
    A_{i_1\cdots i_n }\equiv D_{i_1}\cdots D_{i_n} A,\quad  \bar{A}^{i_1\cdots i_n }\equiv D^{i_1}\cdots D^{i_n} \bar{A}, \quad D^i\equiv K^{i \bar{m}}D_ {\bar{m}} \, ,
\end{equation}
$K^{i \bar{m}}$ is the inverse Kähler metric, and $D_i$ is the K\"ahler covariant derivative: $ D_i W = \partial_i W + W \partial_i K$. The derivative acting on $A$ is $D_i A=\partial_i A =e^K D_i W$, 
$D_iD_j A= D_i \partial_j A = \partial_i\partial_j A-\Gamma^k_{ij}\partial_k A$, etc. 

In  the globally-supersymmetric limit, 
after restoring the factors of $M_P$, it is easy to see that
when $M_P\rightarrow \infty$ 
we have $e^{K/M_P^2}\rightarrow 1$, $A\rightarrow W$, and 
\begin{equation}
     K^{(1)}\rightarrow \frac{\log\mu^2}{32\pi ^2} W_{ij}\bar{W}^{ij} \, .
\end{equation}
Since the K\"ahler covariant derivative reduces to a partial derivative in the globally-supersymmetric limit, $W_{ij}\rightarrow \partial_{\Phi_i}\partial_{\Phi_j}W$, we recover the $\mu$-dependent piece in the globally-supersymmetric expression \eqref{minoneloop}.  There is an extra constant in the parenthesis of \eqref{minoneloop}, which is due to the different regularization schemes. This does not affect the anomalous dimension (arising from the $\mu$-dependent part) or the subsequent running.

As was explained in Section~\ref{sec1}, we focus on the logarithmically-divergent one-loop potential in supergravity  and apply the expression in  \cite{Gaillard:1993es}. We again make use of the non-renormalization theorem to derive the $\beta$-functions, and consequently all the couplings in the one-loop corrected potential will be scale-dependent.

We note that, because the supergravity model includes a dynamical metric and inflation proceeds in an (approximately) de Sitter background, metric renormalization could a priori also contribute to the effective action. Following~\cite{Gaillard:1993es}, the renormalized metric $g_{\mu\nu}^R$ is related to the bare metric by
\begin{equation}
    g_{\mu\nu} =(1-\ep )g_{\mu\nu}^R+\epsilon_{\mu\nu}\,.
\end{equation}
The tensor $\epsilon_{\mu\nu}$ contains terms with derivatives of the scalar field, which induce derivative couplings to the metric. Such couplings lie beyond the scope of this work and will be neglected. The remaining contributions in $\epsilon_{\mu\nu}$ and $\epsilon$ depend on $A$, the tree-level potential, the spacetime Ricci tensor $R_{\mu\nu}$, and the gravitino mass. During inflation, these effects are suppressed by $H^2/ M_P^2\ll 1$, and therefore their contributions to the spacetime Ricci scalar are negligible. This is consistent with the results of  \cite{Ellis:2025bzi}. Equivalently, any such terms can be transferred into the scalar potential via a Weyl rescaling, where they remain similarly suppressed. Thus, in the following we do not consider spacetime curvature effects.

\section{Supergravity Inflation Models with Canonical Kinetic Terms}
\label{new}

We begin by analyzing the radiative corrections in inflation models that are constructed in minimal supergravity, namely, models with canonical kinetic terms, for which $K_{\Phi_i \bar \Phi_j} = \delta_{ij}$. We first consider the simplest case, a single field model introduced by Holman, Ramond and Ross (HRR) \cite{hrr}. We then examine a minimal supergravity realization of the Starobinsky model derived using a method introduced by Kawasaki, Yamaguchi and Yanagida (KYY) \cite{Kawasaki:2000yn} that features an approximate shift symmetry. 

\subsection{New inflation (HRR) model}
\begin{figure}[!ht]
\centering\includegraphics[width=.6\textwidth]{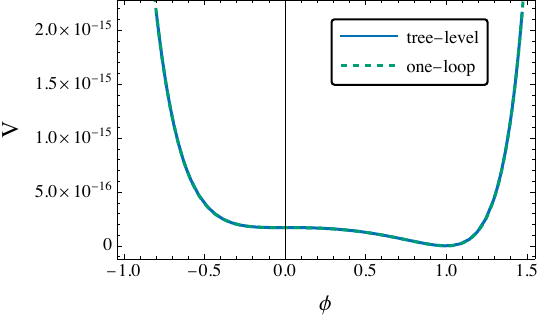}
    \caption{The scalar potential $V(\phi)$ in the HRR model, shown at tree level (blue solid line) and including one-loop corrections (green dashed line).}
    \label{Vnewinf}
\end{figure}

The HRR model \cite{hrr} has a single (complex) scalar superfield with minimal K\"ahler potential and quadratic superpotential. 
At tree-level, the model is defined by
\begin{equation}
    K^{(0)}=\bar{\Phi}\Phi,\quad W=M(\Phi-1)^2 \, ,
\end{equation}
where the mass scale $M \simeq 10^{-8} M_P$ is determined by the amplitude of density perturbations as discussed below.  The tree-level potential along the real $\phi$ direction is shown in Fig.~\ref{Vnewinf} by the solid curve. This is a ``small-field" inflation model, in that
inflation occurs when $\phi \simeq 0$. 
The imaginary part of $\phi$ is always stabilized at Im$\phi=0$ as it has a positive mass squared in that direction. 

A key role in testing this and other inflationary models is played by precision determinations of the tilt, $n_s$, of the spectrum of scalar CMB perturbations. 
In slow-roll models of inflation the tilt is determined to leading order by the parameters
\begin{equation} 
\epsilon \; \equiv \; \frac{1}{2} M_{P}^2 \left( \frac{V'}{V} \right)^2\,, \qquad  \eta \; \equiv \; M_{P}^2 \left( \frac{V''}{V} \right)  \, ,
\label{epsilon}
\end{equation}
where prime refers to the derivative with respect to the inflaton field. 
The tilt is given by
\beq
n_s \;  \simeq \; 1 - 6 \epsilon_* + 2 \eta_* = \; 0.9649 \pm 0.0044 \quad (68\%~{\rm CL}) \, ,
\label{ns}
\eeq
where we have quoted the numerical value from~\cite{Planck}. This constraint is evaluated at $\phi_*$, which is fixed by the pivot scale, $k_* = 0.05~{\rm Mpc}^{-1}$.
There have recently been two new determinations of $n_s$. The ACT DR6 results~\cite{ACT:2025fju} indicate a larger value of $n_s$. When combined with Planck, lensing and DESI DR2 BAO results they yield~\cite{ACT:2025fju}
\begin{equation}
n_s = 0.9752 \pm 0.0030 \quad (68\%~{\rm CL}) \, .
\label{ACT}
\end{equation}
On the other hand, the South Pole Telescope (SPT) Collaboration has provided 3G data~\cite{SPT-3G:2025bzu}
that are consistent with Planck. 
Clearly it would be interesting if radiative corrections were to shift the predictions of Starobinsky-like models to larger values of $n_s$, more compatible with (\ref{ACT}).

Another CMB observable is the tensor-to-scalar perturbation ratio, $r$, which may be expressed in terms of the slow roll parameters as $r \simeq 16 \epsilon_*$. Planck data in combination with observations by BICEP/Keck~\cite{rlimit,Tristram:2021tvh} provide an upper limit to this ratio, $r < 0.036$. In all of the models we consider, the predicted value of $r$ is significantly below this limit. We comment on the impact of radiative corrections on $r$ in the models discussed below. 

The value of $\phi_*$ depends weakly on the reheating temperature, which also sets the number of e-folds between $\phi_*$ and $\phi_{\rm end}$, the field value when inflation ends \cite{LiddleLeach,Martin:2010kz, Ellis:2021kad,Ellis:2025zrf}. Typically, $\phi_* \approx 0.0015$ and $\phi_{\rm end} \simeq 0.68$ (in Planck units), corresponding to $N_*\approx 56$ e-folds of inflation for $\phi > \phi_*$. 
The amplitude of the scalar fluctuations is given in terms of the potential scale $M$ and $\epsilon_*$ by
\beq
A_s \; = \; \frac{V_*}{24 \pi^2 \epsilon_* M_{P}^4 } \simeq 2.1 \times 10^{-9} \, , \label{As} 
\eeq
where $V_* = V(\phi_*) \propto M^2$. 
Using $A_s = 2.1 \times 10^{-9}$ \cite{Planck}, we
obtain $M= 1.3 \times 10^{-8}$ (in Planck units).

For a wide range of reheating temperatures, the HRR model predicts a value of $n_s \approx 0.90 - 0.93$ \cite{Antoniadis:2025pfa}, which is well below the observed value. The tensor-to-scalar ratio, is unobservably small in this model, $r\sim \mathcal{O}(10^{-8})$. However, the discrepancy in $n_s$ can be resolved by a small deformation of the superpotential, taking $W = M(\phi - \lambda)^2$ with $\lambda \ne 1$. For $\lambda \simeq 0.999995$ (with the corresponding shifts in $\phi_*$ and $M$), the value of $n_s$ increases sufficiently to match the Planck or even the ACT determination.

\begin{figure}[!ht]
\centering\includegraphics[width=.65\textwidth]{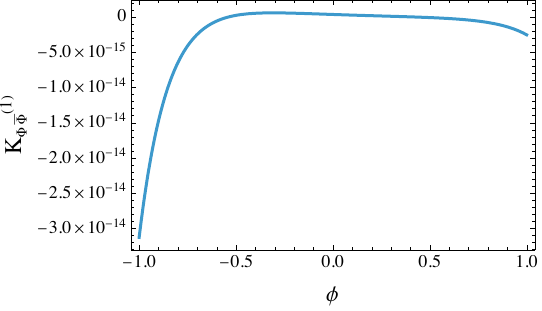}
    \caption{The one-loop corrected Kähler metric as a function of the scalar field $\phi$. We have set $ \mu=M$ with $M=1.3\times 10^{-8}$, and take $\bar{\phi}=\phi$. }
    \label{k1newinf}
\end{figure}

To test the stability of the HRR model, we  introduce the parameters $\lambda_i$ via the superpotential
\begin{equation}
    W=\lambda_1+\lambda_2 \Phi+\lambda_3 \Phi^2\,.
    \label{HRRcouplings}
\end{equation}
The non-renormalization theorem implies that $\lambda_1$ is constant, while the $\beta$-functions for $\lambda_2$, $\lambda_3$ are
\begin{equation}
    \beta_2=\gamma_\phi \lambda_2,\quad     \beta_3=2\gamma_\phi \lambda_3 \, .
\end{equation}
From \eqref{Kgeneral} and fixing $\lambda_i=1$, we obtain
\begin{equation}\begin{aligned}
    K^{(1)}=&\frac{\log\mu^2}{32\pi ^2}  e^{|\Phi|^2}M^2\left[-8 + 12 \bar{\Phi} - 6 \bar{\Phi}^2 + \Phi^4 (-1 + \bar{\Phi})^2 \bar{\Phi}^2 + 
 2 \Phi (6 - 5 \bar{\Phi} + \bar{\Phi}^3) \right.\\&\left.- 
 2 \Phi^3 \bar{\Phi} (-1 + 5 \bar{\Phi} - 
    5 \bar{\Phi}^2 + \bar{\Phi}^3) + \Phi^2 (-6 + 17 \bar{\Phi}^2 - 
    10 \bar{\Phi}^3 + \bar{\Phi}^4)\right] \, ,
\end{aligned}
\end{equation}
from which we may extract $\gamma_\phi$, the anomalous dimension of $\phi$.
The correction to the Kähler metric $K_{\Phi\bar{\Phi}}$ is tiny for $\phi$ in the range $[-1,1]$, as shown in Fig.~\ref{k1newinf}. Therefore, we can treat $\phi$ as a canonically normalized scalar field.
For the initial values of the parameters in Eq.~(\ref{HRRcouplings}), we take $\lambda_2(\mu_0)=-2 M$ and $\lambda_3(\mu_0)= M$, with $M=1.3\times 10^{-8}$, while $\lambda_1 = M$ remains constant. We set the initial renormalization scale to be $\mu_0=1$ and evolve to $\mu=M$, which is of order the Hubble scale at the beginning of inflation.\footnote{Note that the one-loop Kähler metric (and hence correction to the kinetic term) is field-dependent. Therefore, for each inflaton field value  $\phi$, we obtain an anomalous dimension $\gamma_\phi$, and corresponding $\beta$-functions, from which we determine the running couplings $\lambda_i(\mu)$. }

\begin{figure}[!ht]
\centering\includegraphics[width=.6\textwidth]{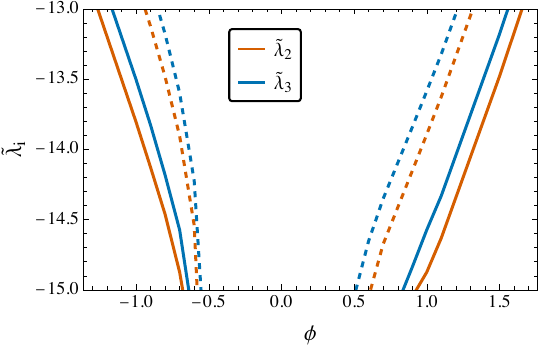}
    \caption{One-loop induced shifts in $\lambda_{2,3}$ as a function of $\phi$. Solid curves: $\lambda=1$, dashed curves: $\lambda = 0.999995$.}
    \label{lamnewinf}
\end{figure}

The running-induced changes in $\lambda_{2,3}$, expressed as $\tilde{\lambda}_2 \equiv \log_{10}(1-\lambda_2/(-2M))$ and $\tilde{\lambda}_3 \equiv \log_{10}(1-\lambda_3/ M )$, are shown in Fig.~\ref{lamnewinf}.
We see that the couplings $\lambda_{2,3}$ run only very weakly, which leads to a negligible difference between the tree-level and one-loop evaluations of the effective scalar potential, as seen in Fig.~\ref{Vnewinf}, where the one-loop-corrected potential, shown by the dashed curve, is barely distinguishable from the tree-level potential shown by the solid curve. Consequently, the one-loop corrections do not significantly affect the predictions of the HRR model for CMB observables, which remain in disagreement with the data. 

In the deformed model with $\lambda = 0.999995$, $\phi_* \sim 7\times 10^{-4}$ 
and $M \simeq 4.3 \times 10^{-8}$, the renormalization-scale boundary conditions are $\lambda_2(\mu_0) = -2 \lambda M$ and $\lambda_3(\mu_0) = \lambda M$. The resulting running of these couplings, shown by the dashed curves in Fig.~\ref{lamnewinf}, is nearly identical to the $\lambda = 1$ case. Consequently, the deformed model continues to fit the data and is essentially unaffected by the one-loop corrections.

\subsection{Models with approximate shift symmetry (KYY)}

The Kähler potential and superpotential in this class of models are
\begin{equation}
K^{(0)}=S\bar{S}-\frac{1}{2}\left(\Phi-\bar{\Phi}\right)^2,\qquad W=MS f(\Phi) \, ,
\end{equation}
where $K^{(0)}$ exhibits a shift symmetry for $\Phi$.
When $S$ is stabilized to zero, the tree-level effective potential is $V = M^2 |f(\Phi)|^2$ \cite{Kawasaki:2000yn, Kallosh:2010ug, Kallosh:2010xz, Kallosh:2011qk}. In the original KYY paper \cite{Kawasaki:2000yn}, the choice $f(\Phi) = \Phi$ was made. This choice yields a quadratic potential $\propto |\Phi|^2$, corresponding to chaotic inflation \cite{Linde:1983gd}, whose predictions are incompatible with current CMB measurements.
The radiative corrections for this model are very small and do not change this conclusion. 

More generally, the KYY construction can realize any potential that can be written as a perfect square. In particular, choosing $f(\Phi) = \sqrt{\frac34} \left(1-e^{-\sqrt{2/3}\lambda\Phi}\right)$ with $\lambda=1$ gives, in terms of $\phi = \text{Re}\Phi$, the Starobinsky potential (\ref{staropot}).
In contrast to the ``new" inflation model of HRR, the Starobinsky model is a ``large" field inflation model. In this case, the last $N_* \approx 55$ efolds are obtained for $\phi_* \simeq 5.35$ with $\phi_{\rm end} \simeq 0.6$  and $M= 1.3 \times 10^{-5}$, yielding $n_s \simeq 0.965$, in good agreement with the Planck data.

\begin{figure}[!ht]
\centering\includegraphics[width=.65\textwidth]{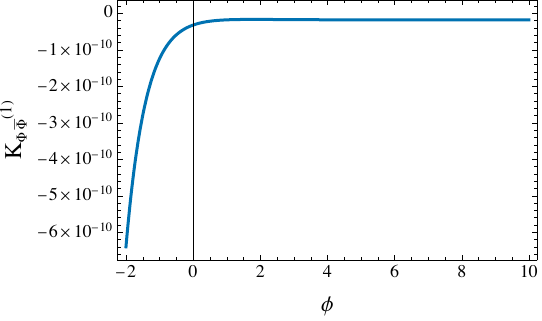}
    \caption{The one-loop K{\"a}hler metric $K^{(1)}_{\Phi\bar{\Phi}}$ as a function of $\phi$. We fix $M=1.3\times 10^{-5}$, and $\lambda=1$ and set $S=0$ and $\text{Im}\,\phi=0$.}
    \label{kkl}
\end{figure}

The complete expression for the one-loop correction to the Kähler potential in the KYY model is given in (\ref{KYYK1}). The one-loop Kähler metric $K^{(1)}_{\Phi\bar{\Phi}}$ is shown in Fig.~\ref{kkl}. For  positive $\phi$, the one-loop correction to the Kähler metric is of order $10^{-11}\ll1$. Similarly, for $\phi>0$ we find $K^{(1)}_{S\bar{S}}\sim 10^{-13}$, which is also negligible compared to its tree-level value. 
The off-diagonal components of the one-loop Kähler metric satisfy $K^{(1)}_{\Phi\bar{S}}=K^{(1)}_{S\bar{\Phi}}$, and vanish at the stabilized values. 

\begin{figure}[!ht]
\centering\includegraphics[width=.6\textwidth]{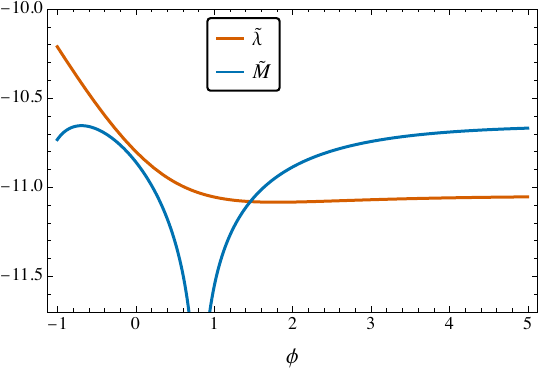}
    \caption{The coupling variations $\tilde{\lambda} $, $\tilde{M}$ as functions of $\phi$.}
    \label{lamkl}
\end{figure}

Therefore, the $\beta$-functions for the running couplings $M,\lambda$ are
\begin{equation}
       \beta_M=M\gamma_S,\quad \beta_\lambda=\lambda\gamma_\phi\,,
   \end{equation}
with \begin{equation}
       \gamma_S=\frac{1}{2}\frac{\partial K^{(1)}_{S\bar{S}} }{\partial \log \mu},\qquad \gamma_\phi=\frac{1}{2}\frac{\partial K^{(1)}_{\Phi\bar{\Phi}} }{\partial \log \mu} \, .
   \end{equation}
At the initial scale $\mu_0=1$ we impose $M(\mu_0) =1.3\times 10^{-5} $ and $\lambda(\mu_0)=1$. We parametrize the deviations of the running couplings from their initial values as $\tilde{\lambda}= \log_{10}(1-\lambda)$, $\tilde{M}= \log_{10}\left|1-\frac{M}{1.3\times 10^{-5}}\right|$, and their values as functions of $\phi=\text{Re}\Phi$ are shown in Fig.~\ref{lamkl}. 

Using the results of \cite{Gaillard:1993es,Gaillard:1996hs}, we find that the one-loop correction to the scalar potential is simply
\begin{equation}
    V_1=\frac{\log\mu^2}{4\pi^2}M^4\left(1-e^{-\sqrt{2/3}\lambda\phi}\right)^4 \,.
\end{equation}
This contribution is negligible compared to the tree-level potential. Consequently, the Starobinsky realization of the KYY model remains compatible with the CMB data.

\section{No-scale supergravity}
\label{nssugra}

Minimal supergravity models are defined as theories on a flat field-space manifold, i.e., with a K\"ahler metric $K_{ij} = \delta_{ij}$. We now consider the class of theories with a maximally symmetric metric, which can be constructed from a K\"ahler potential of the form 
\beq
K \; = \; -3 \, n  \, \ln\left(T + \bar{T} - \frac{|\chi_i|^2}{3}\right) \, ,
\label{n-sK}
\eeq
where $n$ is a constant, $T$ can be associated with a modulus field and the other $N-1$ fields, $\chi_i$, with matter fields~\cite{no-scale1,EKN}. 
This form of the K\"ahler potential has been derived as the low-energy effective theory from string models \cite{Witten}. The field-space manifold has constant Ricci curvature, $R = N(N+1)/3n$ and the fields parameterize an SU($N$,1)/SU($N$)$\times$U(1) coset space. 
The field-space manifold can, equivalently to (\ref{n-sK}), be reparameterized symmetrically by $N$ complex scalar fields, $y_i$
\beq
K \; = \; -3 \, n \, \ln\left(1 - \sum _{i=1}^N\frac{|y_i|^2}{3}\right) \, ,
\label{n-sKy}
\eeq
through a set of field redefinitions. In the absence of a superpotential, the scalar potential vanishes identically $(V=0)$, for all field values, hence the name no-scale \cite{ELNT,no-scale2}.
For simplicity, we set $n=1$ for the models to be discussed below.

\subsection{No-Scale Avatars of the Starobinsky Model}
\label{sec:noscaleavatars}

At least two chiral fields are required to obtain a Starobinsky-like inflation model in no-scale supergravity~\cite{Ellis:2013nxa}. Here we focus on the minimal two-field setup. 
In the $T,\chi$ basis (\ref{n-sK}), it is possible to construct models
where either $T$ or $\chi$ plays the role of the inflaton, provided the other field is stabilized. In the symmetric $y_i$ basis (\ref{n-sKy}) for the two-field case, either $y_1$ or $y_2$ can be used as the inflaton, as long as the other field is stabilized at the origin. After stabilization, the real part of $y_1$ or $y_2$
can be redefined to yield a canonically-normalized inflaton with
\beq
{\rm Re}~y_i = \pm \sqrt{3} \tanh\left(\frac{\phi}{\sqrt{6}} \right)\,,
\label{branches}
\eeq
for $i=1,2$.
The four branches of solutions ($i = 1$ or 2 and the sign of the field redefinition (\ref{branches})) can be related by 
transformations consistent with the $SU(2,1)/SU(2)\times U(1)$ symmetry: $y_1 \to \alpha y_1 + \beta y_2$ and $y_2 \to \beta^* y_1 + \alpha^* y_2$, for arbitrary constants $\alpha,\beta$.
Assuming that $y_2$ is fixed, we can rewrite the Starobinsky potential (\ref{staropot}) in terms of $y_1$ using Eq.~(\ref{branches}),
obtaining 
\beq
V = M^2 \frac{{\rm Re}~y_1^2}{(1+{\rm Re}~y_1/\sqrt{3})^2} \, , \label{match1}
\eeq
which corresponds to 
\begin{equation}
V \; = \; \frac{M^2 |y_1|^2 ~ |1 - y_1/\sqrt{3}|^2}{(1 - |y_1|^2/3)^2} \, ,
\label{V1}
\end{equation}
when written in terms of the complex field, $y_1$. 

The scalar potential derived from Eq.~(\ref{n-sKy}) 
for a general superpotential $W(y_1,y_2)$ takes the form
\begin{equation}
    V = \frac{1}{(1 - |y_1|^2/3)^2} \left( (1 - |y_1|^2/3) |W_1|^2 + |W_2|^2 - 3 |W|^2 +(y_1 W_1 W^* + {\rm h.c.}) \right)\, ,
\label{useful}
\end{equation}
where $W_{1,2} = \partial W/\partial y_{1,2}$, and we have assumed $\langle y_2 \rangle = 0$ and $n=1$.

In the notation of \cite{Ellis:2018zya}, starting with Branch I (where the inflaton is associated with $y_1$, $\langle y_2 \rangle = 0$, and the $+$ sign is chosen in (\ref{branches})), 
we can write a general superpotential of the form
\beq
W(y_1, y_2)= M\left(a y_1+b y_1^2+c y_1^3+d y_2+e y_2 y_1+f y_2 y_1^2 + g(y_1, y_2)\right) \, , 
\label{genW}
\eeq
with $a,b,c,d,e,f$ arbitrary coefficients, $g(y_1, 0)
= 0$, $\partial g/ \partial y_1 (y_1 , 0)$  $= 0$ and $\partial g/ \partial y_2 (y_1, 0) = 0$.
The function $g(y_1,y_2)$ may also include terms containing factors $y_2^m$, as these would not contribute to $V$, since $\langle y_2 \rangle = 0$. The other branches can be obtained from Branch I with the choices $\alpha = -1$, $\beta = 0$ or $\alpha = 0$, $\beta = 1$ \cite{Ellis:2018zya,building}. By matching Eq.~(\ref{V1}) with Eq.~(\ref{useful}), and using the general superpotential in Eq.~(\ref{genW}) for $W$, we obtain a solution for the couplings in (\ref{genW})
\beq
a = d = 0, \qquad c = -\frac{b \left(\sqrt{1-4 b^2}+2\right)}{3 \sqrt{3}}, \qquad e = \pm \sqrt{1-4 b^2},\qquad  f = \mp \frac{\sqrt{1-4 b^2} - 2 b^2}{\sqrt{3}} \, ,
\label{solcoef}
\eeq
parameterized by $b$, all of which yield the Starobinsky potential (\ref{staropot}) (with other solutions given in \cite{Ellis:2018zya}). Before moving to the calculation of radiative corrections in specific no-scale realizations of the Starobinsky potential, we summarize a few general observations applicable to the entire class of no-scale models considered below.

\subsection{The gravitino mass}

The loop correction to the supergravity scalar potential involves complicated expressions and running couplings. However, when a no-scale model gives the Starobinsky potential at tree-level, we can obtain a qualitative understanding of the loop correction by considering the gravitino mass. We recall that in supergravity the scalar potential is given by
\begin{equation}
    V=F^2-3m_{3/2}^2 \quad \text{with}\quad  F=\sqrt{e^KD_iWD_{\bar{j}}\bar{W}K^{i\bar{j}}} \quad {\rm and} \quad m_{3/2}^2 = e^K |W|^2 \, ,
\end{equation}
where $F$ quantifies the amount of supersymmetry breaking and is related to the mass splitting in the particle spectrum. When supersymmetry is unbroken, the CW potential vanishes when evaluated using the supertrace formula. Once supersymmetry is broken, the resulting mass splittings generate a non-zero one-loop correction. In principle, if supersymmetry is badly broken ($F$ is large), then the one-loop correction would also be large.

In no-scale models that yield the Starobinsky potential at tree-level the potential is flat at large field values. As a consequence, if $m_{3/2}$ grows at large field values, so also does $F$, leading to large quantum corrections. In order to illustrate this point, we consider the no-scale Kähler potential in the $y$ basis given in Eq.~(\ref{n-sKy}) with $n = 1$. 
When one of $\{y_1,y_2\}$ is stabilized to zero, the other is related to the canonical inflaton field by 
Eq.~(\ref{branches}). 
As noted in Section~\ref{sec:noscaleavatars}, different possibilities give rise to four branches \cite{Ellis:2018zya}, but they all yield the relation
\begin{equation}
  \left<  e^{K/2}\right>=\cosh\left(\frac{\phi}{\sqrt{6}}\right)^3 \, ,
\end{equation}
which increases exponentially for large $\phi$, where $\left<\cdot\right>$ denotes evaluation at the stabilized field value.
For $m_{3/2}$ not to grow at large $\phi$, $W$ should either decrease exponentially, or vanish identically, when one of the $y$ fields is stabilized. However, if $W$ is given by a polynomial of $y_i$ with positive exponents, as in (\ref{genW}), $W$ will never decrease exponentially. Therefore, $W$ must vanish when evaluated at the stabilized background values. Assuming, without loss of generality that $\left<y_2\right>=0$, $W$ can then be written as
\begin{equation}
    W=y_2\,u(y_1,y_2) \, , \label{wy1}
\end{equation}
where $u$ is a sum of polynomials. Note that Eq.~\eqref{wy1} corresponds to \eqref{genW} with $a=b=c=0$.
With $\left<y_2\right>=0$, this structure ensures that the gravitino mass vanishes and we need only require that the supersymmetry-breaking parameter $F$ stays approximately constant ($F=\sqrt{V}\simeq M$) to maintain a plateau at large field values.
On the other hand, if $W$ cannot be written in the form \eqref{wy1}, $m_{3/2}$ (and hence $F$) grows exponentially at large $\phi$, i.e., the supersymmetry-breaking becomes increasingly worse, and the one-loop corrections to the potential grow accordingly.

Up to field redefinitions, we can always choose $y_1$ from Eq.~(\ref{branches}) with $\left<y_2\right>=0$,  and the Starobinsky potential in this basis is given by Eq.~\eqref{match1}.
From \eqref{wy1}, we also have $\left<D_{y_2}W\right>=\left<u\right>$, $\left<D_{y_1}W\right>=0 $, so the scalar potential is 
\begin{equation}
    V=\frac{9}{(y_1^2-3)^2}\left<u^2\right> \, . \label{match2}
\end{equation}
Matching \eqref{match1} and \eqref{match2}, we find that
\begin{equation}
    u(y_1,y_2)=\frac{1}{\sqrt{3}}My_1\left(y_1-\sqrt{3}\right)+h(y_1,y_2) \, ,
\end{equation}
where the function $h$ satisfies $h(y_1,y_2=0)=0$. If $h$ is identically zero, we recover the Cecotti model \cite{Cecotti}, to be discussed in Section~\ref{sec:cecotti}. Restricting to a renormalizable superpotential, i.e., a polynomial that is at most cubic, we can parameterize $h$ as $h(y_1,y_2)= M\left(a^\prime y_2^2-b^\prime  y_1 y_2+c ^\prime y_2\right)$ where $a^\prime, b^\prime, c^\prime$ are arbitrary coefficients. The superpotential can then be written as 
\begin{equation}
    W=M\left(\frac{1}{\sqrt{3}} y_1^2y_2-y_1y_2+a^\prime y_2^3-b ^\prime y_1 y_2^2+c ^\prime y_2^2\right) \, . \label{Wm320}
\end{equation}
Hence, with this superpotential form, the one-loop correction to the scalar potential remains small up to some very large inflaton field value. Furthermore, 
as we will see next, the superpotential form \eqref{Wm320} also guarantees the cancellation of the leading singularities in the one-loop K\"ahler potential.

\subsection{Singularities in the Kähler potential}
\label{sec53}

In no-scale supergravity, the tree-level Kähler potential has a logarithmic singularity, while the tree-level Kähler metric
has a quadratic singularity. More generally, the importance of the 
one-loop correction is correlated with its degree of singularity.
To facilitate its discussion, we work in the $\{y_1,y_2\}$ basis, in which the four model branches are related by sign flips and/or the interchange $y_1\leftrightarrow y_2$. We then focus on the branch defined by $y_1=\sqrt{3}\tanh\left(\frac{\phi}{\sqrt{6}}\right)$ with $\left<y_2\right>=0$.

The tree-level Kähler potential \eqref{n-sKy} diverges logarithmically as $y_1\rightarrow \sqrt{3}$ or as the canonically normalized field $\phi\rightarrow+\infty$. Since the one-loop Kähler potential \eqref{Kgeneral} has factors of $e^K$, one would expect it to diverge as $1/( y_1-\sqrt{3})^k$ for $k \ge 1$, modulo the possibility of special cancellations that we discuss later. Consequently, at some large field value $\phi$, the one-loop Kähler potential could dominate over the tree-level contribution, resulting in  a large deformation of the potential and a breakdown of perturbation theory.
For successful inflation, either $W$ must take a special form so the one-loop Kähler potential has no singularity, or the one-loop Kähler potential must remain sub-dominant up to field values $\phi$ far beyond the CMB scale, $\phi_*$, so that the resulting deformation does not affect the CMB observables. We next discuss these two possibilities.\\

\noindent
{\bf Case I: No singularity in $K^{(1)}$}

\vspace{2mm}
\noindent
We start with the general form of the superpotential \eqref{genW} satisfying \eqref{solcoef}, again only including terms up to cubic order in $g(y_1,y_2)$. Using \eqref{Kgeneral}, we obtain the one-loop Kähler potential evaluated at $\left<y_2\right>=0$. Requiring that $K^{(1)}$ is non-singular at $y_1=\sqrt{3}$, we obtain the following general form for the superpotential
 \begin{equation}
    W=M\left(\frac{1}{\sqrt{3}}y_1^2y_2- y_1y_2+a^\prime y_2^3-b^\prime  y_1y_2^2  +\sqrt{3}b^\prime  y_2^2\right) \, . \label{Wnosing}
\end{equation}
Since $K^{(1)}$ is non-singular, the most divergent terms in the Kähler metric $K_{y_1\bar{y}_1}$ are
\begin{equation}
K^{(0)}_{y_1\bar{y}_1} \sim  \frac{3}{4(y_1-\sqrt{3})^2},\quad 
K^{(1)}_    {y_1\bar{y}_1} \sim  \frac{M^2}{4(y_1-\sqrt{3})^2} \, .
\end{equation}
In addition, since $M^2\sim 10^{-10}$, the one-loop Kähler metric is always subdominant, and therefore the one-loop corrections to the couplings stay small.
We note that \eqref{Wnosing} is a subcase of \eqref{Wm320} with $c^\prime=\sqrt{3}b^\prime $, and the Cecotti model \cite{Cecotti} is a special example of \eqref{Wnosing} with $a^\prime =b^\prime =0$.\\

\vspace{2mm}

\noindent
{\bf Case II: singular $K^{(1)}$, but subdominant up to large $\phi$}

\vspace{2mm}
\noindent
We now relax the no-singularity condition, and instead impose the weaker requirement that $K^{(1)}_{y_1\bar{y}_1} \ll K^{(0)}_{y_1\bar{y}_1} $ up to some field value $\phi > \phi_*$. Performing a Laurent expansion of $K^{(1)}_{y_1\bar{y}_1}$ about $y_1=\sqrt{3}$, and assuming that the non-vanishing coefficients in the superpotential are of order one, gives
\begin{equation}
    K^{(1)}_{y_1\bar{y}_1} \simeq M^2
    \left[\frac{c_5}{(y_1-\sqrt{3})^5}+\frac{c_4}{(y_1-\sqrt{3})^4}+\frac{c_3}{(y_1-\sqrt{3})^3}+\frac{c_2}{(y_1-\sqrt{3})^2}+\frac{c_1}{(y_1-\sqrt{3})}+c_0
    \right] \, ,
    \label{eq:laurent}
\end{equation}
where the coefficients $c_n$ are expected to be order one and the leading singularity is of fifth order.
For small $\phi$ ($y_1\rightarrow0$), the one-loop contribution is safely subdominant because $  K^{(1)}_    {y_1\bar{y}_1} \sim 10^{-10}$, whereas $ K^{(0)}_{y_1\bar{y}_1}$ is of order one. For large $\phi$ ($y_1\rightarrow \sqrt{3}$), the first three terms in \eqref{eq:laurent} will eventually dominate, if they are present. If we impose the condition that the terms containing ${\cal O}((y_1-\sqrt{3})^{-5}, (y_1-\sqrt{3})^{-4}
, (y_1-\sqrt{3})^{-3})$  all vanish, we recover the previous solution for $W$, namely \eqref{Wnosing}. 
However, if we retain the $(y_1-\sqrt{3})^{-3}$ term but assume that the ${\cal O}((y_1-\sqrt{3})^{-5}, (y_1-\sqrt{3})^{-4})$ terms vanish, we obtain a more general solution
\begin{equation}
    W=M\left(\frac{1}{\sqrt{3}}y_1^2y_2- y_1y_2+a^\prime y_2^3-b^\prime  y_1y_2^2  +c^\prime  y_2^2\right) \, .
\end{equation}
which is exactly \eqref{Wm320}. Namely, imposing  $m_{3/2}=0$ or imposing that the ${\cal O}((y_1-\sqrt{3})^{-5}, (y_1-\sqrt{3})^{-4})$ terms in $K^{(1)}_{y_1\bar{y}_1}$ vanish, both lead to the same solution.

If the ${\cal O}((y_1-\sqrt{3})^{-3})$ term does not vanish, the metric will remain subdominant so long as $(y_1 - \sqrt{3}) \lesssim 10^{-10}$ which corresponds to $\phi \sim 30$. If $W$ does not take the above form, e.g., see \eqref{wzybasis} below, then the ${\cal O}((y_1-\sqrt{3})^{-5}, (y_1-\sqrt{3})^{-4})$ terms in the Kähler metric are non-vanishing. It may still be possible that $ K^{(1)}_{y_1\bar{y}_1} $ stays subdominant up to some large field value, but that will depend on the details of the coefficients.

\section{One-loop corrections to specific avatars} 
\label{oneloopav}

We are now in a position to examine specific realizations of the Starobinsky model within no-scale supergravity. As discussed in Section~\ref{nssugra}, the tree level theory admits a continuous family of superpotentials, characterized by the general form in Eq.~(\ref{genW}) together with the coefficient relations given in Eq.~(\ref{solcoef}). In addition, further solutions are possible using the no-scale symmetry \cite{Ellis:2018zya}. At tree level, all such superpotential solutions reproduce the Starobinsky scalar potential along the real trajectory of one of the two complex scalar fields, after transforming to a canonically normalized inflaton. However, beyond tree level, these avatars need not be equally well-behaved. To avoid introducing one-loop divergences,
we must impose restrictions on the possible solutions for the superpotential, making some avatars safer than others.

\subsection{Wess-Zumino model} 

A Starobinsky-like  model can be constructed very simply in no-scale supergravity by choosing a Wess-Zumino superpotential for a matter-like field $\chi$. In this case, the tree-level  K\"ahler potential in Eq.~(\ref{n-sK}) (with $n=1)$ is supplemented by the superpotential \cite{eno6}
 \begin{equation}
W = M \left(\frac{1}{2}\chi^2 -\frac{1}{3\sqrt{3}}\chi^3\right) \,, 
\label{modelWZ}
\end{equation}
where, in principle, the quadratic and cubic terms carry independent couplings that run. To make this explicit, we introduce a separate coupling $\lambda$ for the cubic term and allow both $M$ and $\lambda$ to run, so that the superpotential is rewritten as
\begin{equation}
    W=  \frac{1}{2}M\chi^2 -\frac{1}{3\sqrt{3}}M\lambda\chi^3 \, . 
\end{equation}

This model can be re-expressed in the $y_{1,2}$ basis by using Eq.~(\ref{n-sKy}) for the tree-level K\"ahler potential together with the field redefinitions
\begin{equation}
T= \frac{1}{2} \left( \frac{1 - {y_2}/{\sqrt{3}}}{1 + {y_2}/{\sqrt{3}}} \right)\,, \qquad \;
\chi = \frac{y_1}{1 + {y_2}/{\sqrt{3}}} \, ,
\label{Tphirewrite}
\end{equation}
and the corresponding inverse transformations
\begin{equation}
y_1 = \frac{2 \chi}{1 + 2 T}\,, \qquad y_2 = \sqrt{3} \left( \frac{1 - 2 T}{1 + 2 T} \right) \, .
\label{Tphiwrite}
\end{equation}
Under the change of field variables in (\ref{Tphirewrite}) and (\ref{Tphiwrite}), the
effective superpotential is transformed to
\begin{equation}
W(T, \phi) \; \to \; \left( 1 + {y_2}/{\sqrt{3}} \right)^3 W(y_1,y_2)  \, .
\label{Wtilde}
\end{equation}
With this change of basis, the Wess-Zumino superpotential (\ref{modelWZ}) becomes 
\begin{equation}
    W=M \left(\frac{ y_1^2}{2} -   \frac{y_1^3}{3 
     \sqrt{3}}  + \frac{y_1 ^2 y_2}{2 \sqrt{3}}   \right)\,.
     \label{wzybasis}
\end{equation}
In this form, this superpotential corresponds to the general solution (\ref{solcoef}) with $b=1/2$, $c=-1/3\sqrt{3}$, and $f = 1/2\sqrt{3}$. 
With ${\rm Re} \ y_1=\sqrt{3}\tanh (\phi/\sqrt{6})$, and restricting to the real trajectory ${\rm Im}\  y_{1,2}=0$, we recover the Starobinsky potential in Eq.~(\ref{staropot}).
This superpotential does not belong to the radiatively safe class of solutions identified in Section~\ref{sec53}, so we expect that the one-loop correction becomes significant at large field values.
(A complete expression for the one-loop K{\" a}hler potential in this model is given in (\ref{WZK1}) {\it et seq.}) However, since the CMB observables are very sensitive to the fine-tuning of model parameters \cite{Antoniadis:2025pfa}, we need to account carefully for all relevant effects, such as the running of the couplings and the stabilization of $y_2$, in order to determine the validity of the no-scale Wess-Zumino model.

As noted above, obtaining the Starobinsky potential requires stabilizing one of the two complex fields (either $T$ or $y_2$ in this formulation). 
A specific stabilization mechanism for $y_2$ and the imaginary part of $y_1$ is obtained by adding a quartic term to the Kähler potential \cite{Ellis:1984bs,Ellis:2013nxa,Ellis:2015kqa,building}
\begin{equation}
    K=-3\log\left(1-\frac{|y_1|^2}{3}-\frac{|y_2|^2}{3}+\frac{|y_2|^4}{\Lambda^2}\right) \, ,\label{stabK}
\end{equation}
where $\Lambda\lesssim1$ sets the stabilization scale.  We also include a quadratic term in the superpotential and introduce running couplings $\lambda_1,\lambda_2,\lambda_3, c^\prime$:
\begin{equation}
   W=M \left(\lambda_1\frac{ y_1^2}{2} -  \lambda_2 \frac{y_1^3}{3 
     \sqrt{3}}  + \lambda_3\frac{y_1 ^2 y_2}{2 \sqrt{3}} +c^\prime y_2^2  \right) \, . \label{stabW}
\end{equation}
The $c^\prime$ term is an explicit example of a contribution of the type $g$ introduced in \eqref{genW}. Note that with the modified Kähler potential in Eq.~\eqref{stabK}, the effective field theory is defined only up to the scale $\Lambda$, which may lie below the Planck scale (though still above the inflationary scale). Accordingly, the boundary conditions for the running couplings are imposed at $\Lambda$.

\begin{figure}[ht]
\centering\includegraphics[width=.5\textwidth]{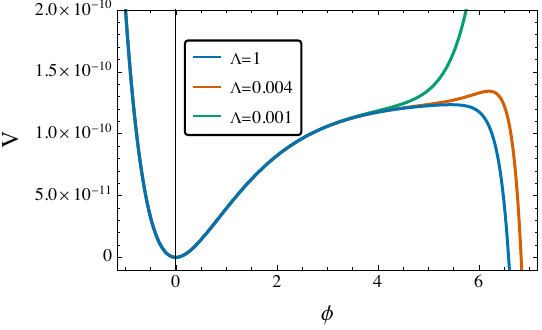}\includegraphics[width=.5\textwidth]{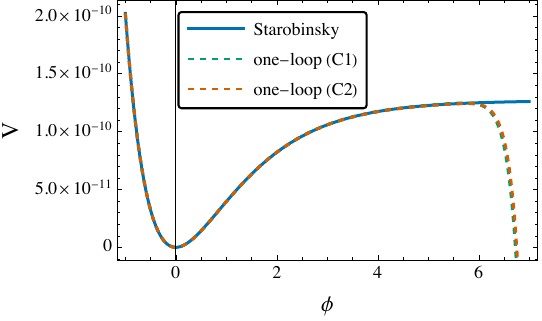}
    \caption{Left panel: The one-loop corrected inflaton potential for different values of $\Lambda$. Right panel: The one-loop corrected potential compared with the Starobinsky potential \eqref{staropot} as a function of the canonical scalar field $\phi$, with $\mu=M=1.3\times 10^{-5}$. Green dashed curve: one-loop corrected potential with the first set of conditions (C1); red dashed curve: one-loop corrected potential with the second set of conditions (C2).}
    \label{stab1}
\end{figure}

The stabilized value of $y_2$ is determined from \eqref{stabK} and \eqref{stabW}. At tree level, the quartic term in the K\"ahler potential together with the quadratic $y_2$ term in the superpotential leaves the scalar potential along the real $y_1$ direction unchanged, reproducing the Starobinsky model (after the field redefinition to the canonical field $\phi$). However, at one loop, these terms can significantly affect the scalar potential. This is illustrated in the left panel of  Fig.~\ref{stab1} which shows the one-loop potential for $\Lambda = 1, 0.004$ and 0.001. We emphasize that at large field values ($\phi \gtrsim 6$), the one-loop potential is no longer quantitatively reliable, since two-loop contributions are expected to become non-negligible. By contrast, the coupling $c^\prime$ does not greatly affect the potential during inflation, but does determine the mass of $y_2$ at the minimum. 

To determine the running couplings at each field value, we note that for a superpotential of the form $W=L^ i y_i +M^{ij }y_{i}y_j+Y^{ijk}y_iy_jy_k$, the $\beta$-functions are given by
 \begin{equation}
    \begin{aligned}
        &\beta_{L^i}=\gamma^i {}_n L^n \, ,
        \\  &\beta_{M^{ij}}=\gamma^i {}_n M^{nj}+\gamma^j {}_n M^{in} \, ,
        \\   &\beta_{Y^{ijk}}=\gamma^i {}_n Y^{njk}+\gamma^j {}_n Y^{ink}+\gamma^k {}_n Y^{ijn} \, .
    \end{aligned}\label{betafunc}
\end{equation}
As specific examples, we show in the right panel of Fig.~\ref{stab1} the one-loop scalar potential for two sets of initial conditions: (C1)  with $\Lambda=0.009$,  $c^\prime =0.01$ and $\lambda_i(\mu_0) = 1$ and (C2) with $\Lambda=0.006$,  $c^\prime (\mu_0)=0.01$, $\lambda_1(\mu_0)=1 + 10^{-5},\lambda_2(\mu_0)=1 +3\times  10^{-5},\lambda_3(\mu_0)=1$, and fixing the initial scale $\mu_0=\Lambda$. The corresponding potentials are shown by the red and green dashed curves in the right panel of Fig.~\ref{stab1}. These are compared with the Starobinsky potential with $M=1.3\times 10^{-5}$. For each $\phi$ we solve and use the precise value of $\left<y_2\right> \approx 0$. For (C1),  we obtain $\left<y_2\right>\sim 10^{-3}$ for small $\phi$, and it becomes closer to zero for larger field values, ensuring a Starobinsky-like plateau during inflation over a limited range in $\phi$.
At the minimum, $\left<y_2\right>=0$.

\begin{figure}[ht]
\centering\includegraphics[width=.5\textwidth]{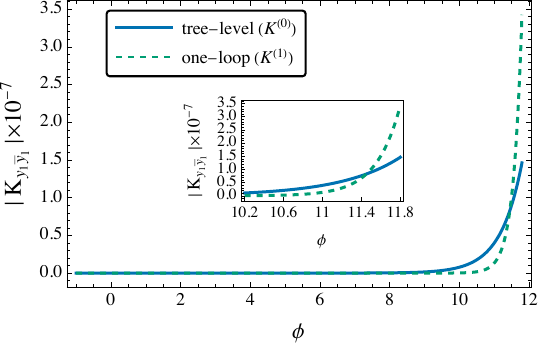}\includegraphics[width=.5\textwidth]{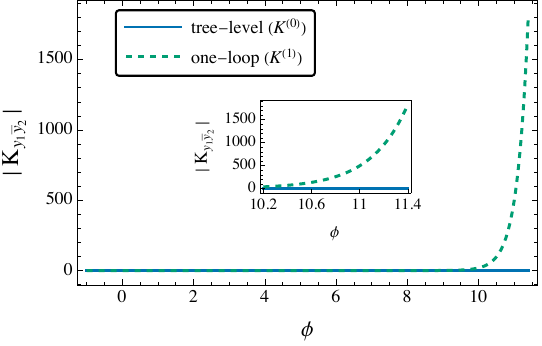}
    \caption{The one-loop Kähler metric in the no-scale Wess-Zumino model as a function of the canonical scalar field $\phi$. Left: the diagonal $y_1 {\bar y}_1$ component; right: the off-diagonal $y_1 {\bar y}_2$ component. We fix $\mu=M=1.3\times 10^{-5}$ and $\lambda_i=1$, $\Lambda=0.009$, $c^\prime=0.01$ (C1). The corresponding components for (C2) are qualitatively similar.}
    \label{kcompWZ}
\end{figure}

The Kähler metric for the (C1) values of $\Lambda$ and $c^\prime$ is shown in Fig.~\ref{kcompWZ}. In the left panel, we show the $y_1 {\bar y}_1$ component. Up to  $\phi\sim8$, which is of interest for CMB predictions, the one-loop Kähler metric is always subdominant. Thus $\phi$ (determined from $y_1$) is a good approximation to the canonical scalar. The right panel shows the off-diagonal component of the K\"ahler metric $K_{y_1\bar{y}_2}$. For large field values $\phi\gtrsim 6$, we have $\left<y_2\right>\simeq  0$ thus $K_{y_1\bar{y}_2}\simeq 0$ at tree-level, whereas the one-loop correction becomes dominant and contributes significantly to the scalar potential. The metric for the conditions (C2) is very similar. 

\begin{figure}[ht]
\centering\includegraphics[width=.6\textwidth]{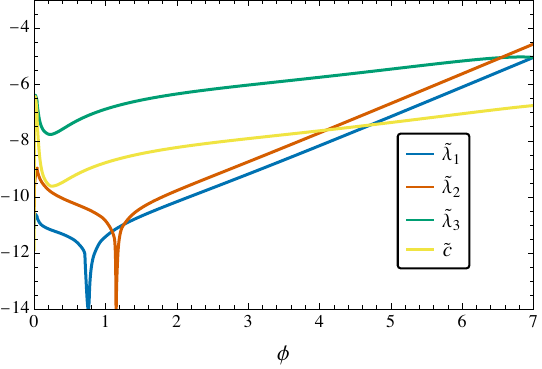}
    \caption{Components of the one-loop corrections to the $\lambda_i$ and $c^\prime$ as functions of the canonical scalar field value $\phi$ in the no-scale Wess-Zumino model. We fix $\mu=M_0=1.3\times 10^{-5}$ and take the initial conditions $\lambda_i(\mu_0)=1$, $c^\prime(\mu_0)=0.01$,  corresponding to (C1). The running obtained from the (C2) initial conditions is qualitatively similar.}
    \label{lamcompWZ}
\end{figure}

Fixing $\lambda_1(\mu_0)=\lambda_2(\mu_0)=\lambda_3(\mu_0)=1$ and $c^\prime (\mu_0)=0.01$ (C1), we solve the $\beta$-functions for each field value. The deviations of the couplings from their initial value are defined as $\tilde{\lambda}_i\equiv \log_{10}|1-\lambda_i|$, $\tilde{c}\equiv \log_{10}|0.01-c^\prime|$, and are shown in Fig.~\ref{lamcompWZ}. The running of the couplings using (C2) is similar. 

The one-loop correction to the scalar potential becomes large at $\phi\sim 6$. At this field value, the  gravitino mass is of order $100M$, so the supersymmetry breaking scale $F \sim m_{3/2} M_P > \Lambda$, indicating a breakdown of our effective treatment of stabilization \cite{Dudas:2017kfz}.
For the (C1) initial conditions, we obtain $n_s \simeq 0.934$, which is significantly outside the CMB range. However, as the one-loop correction becomes important around $\phi_*$, this prediction is also highly sensitive to tuning of the tree-level superpotential couplings \cite{Antoniadis:2025pfa}.
By contrast, for the same parameter set, but using the (C2) initial conditions, we find instead $n_s=0.964$, $r=0.004$, $\phi_*=5.4$, $N_*=55$, which is consistent with the CMB data.

We can therefore conclude that one-loop corrections in the no-scale Wess-Zumino realization of the Starobinsky model are very important. This can be understood from the fact that supersymmetry is badly broken during inflation since both $F$-terms ($F_{y_1}$ and $F_{y_2}$) are large compared with the inflationary scale. Consequently, the protection afforded by supersymmetry with respect to radiative corrections is absent. Nevertheless, the model is still viable, provided some tuning in the tree-level parameters is imposed to keep the inflaton potential sufficiently close to the Starobinsky form and thereby preserve its successful predictions for $n_s$ (and $r$).

\subsection{Cecotti model}
\label{sec:cecotti}

The Starobinsky-like inflationary realization  known as the Cecotti model~\cite{Cecotti} is also contained in the general class of solutions in Eq.~(\ref{solcoef}). Choosing  $b=0$ implies $c=0$, $e=-1$ and $f=1/\sqrt{3}$, giving the superpotential
\beq
W=M\left(-1+\frac{1}{\sqrt{3}} y_1\right)y_1 y_2 \, .
\label{Cy1y2}
\eeq
This belongs to the class of models in which supersymmetry remains under control during inflation. That is, the superpotential takes the form given in Eqs.~(\ref{Wm320}) and (\ref{Wnosing}).  For the inflationary trajectory with $\langle y_2 \rangle = 0$, the gravitino mass vanishes and the one-loop K\"ahler correction is either nonsingular or remains subdominant even at large inflaton values. 
Using the field redefinitions in Eq.~(\ref{Tphiwrite}), the model may be rewritten in the $T,\chi$ basis as \cite{Ellis:2013nxa}
\begin{equation}
W \; = \; M \chi \left(T-\frac12\right)\left[\sqrt{3} \left(T + \frac12 \right) -  \chi \right] \, .
\label{W4}
\end{equation}
The stabilization of $y_2$ corresponds to fixing $T$ and associating the inflaton with $\chi$. After the  field redefinition $\chi = \sqrt{3}\tanh(\phi/\sqrt{6})$ we recover the Starobinsky potential. 
Note that, in this case, the addition of the quartic term in the K\"ahler potential fixes $y_2 = 0$ for all inflaton field values. 

By transforming $y_1\to -y_2$ and $y_2 \to -y_1$ ($\alpha = 0$ and $\beta = -1$) in Eq.~(\ref{Cy1y2}), we recover the more familiar form for the Cecotti superpotential
\begin{equation}       
W=\sqrt{3} M \chi \left(T - \frac12\right) \, .
\end{equation}
In this basis, with $\langle y_1 \rangle = 0$ fixed, we impose $\left<\chi\right>=0$, and identify the inflaton with the real part of the modulus field via ${\rm Re}\,T=\frac{1}{2}e^{\sqrt{\frac{2}{3}}\phi}$. Stabilization can be achieved by adding a quartic term to the K\"ahler potential, as in the Wess-Zumino realization given in Eq.~(\ref{stabK}). 
   
In order to study the effects of radiative corrections in this model, we write the superpotential as 
   \begin{equation}
W=M\left(-\lambda _1 +\frac{\lambda_2}{\sqrt{3}}  y_1\right)y_1 y_2  \, ,
       \label{cecgen}
   \end{equation}
where both $\lambda_1$ and $\lambda_2$ are running parameters, with the  boundary conditions $\lambda_1(\mu_0)=\lambda_2(\mu_0)=1$, and we fix  $M=1.3\times 10^{-5}$.

\begin{figure}[ht]
\centering\includegraphics[width=.6\textwidth]{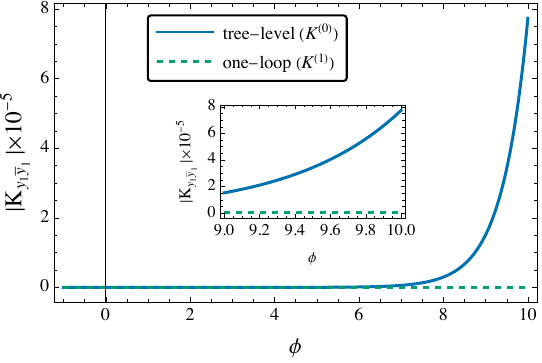}
    \caption{Comparison of the tree-level and one-loop contributions to the $y_1 {\bar y}_1$ component of the Kähler metric in the Cecotti model~\cite{Cecotti}, as a function of the canonically normalized inflaton field $\phi$. When $y_2$ is stabilized to $0$, the off-diagonal component vanishes identically at both tree level and one loop.}
    \label{kcompc}
\end{figure}

The Kähler metric (evaluated for $\left<y_2\right>=0$) is shown  in Fig.~\ref{kcompc}. The one-loop contribution is \textit{always} much smaller than the tree-level term, consistent with the absence of a large-field singularity in the one-loop Kähler potential discussed in Section~\ref{sec53}. The complete one-loop Kähler potential is given in (\ref{CecottiK1}). We note that the off-diagonal $K_{y_1\bar{y}_2}$ is identically zero at both tree level and one loop.

\begin{figure}[ht]
\centering\includegraphics[width=.6\textwidth]{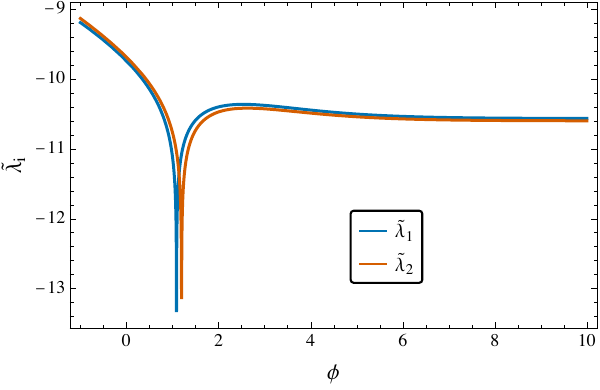}
    \caption{Deviations of the one-loop  corrected couplings $\lambda_1(\mu)$, $\lambda_2(\mu)$ from their initial values in the Cecotti model~\cite{Cecotti}, shown as functions of the canonically normalized inflaton field $\phi$.}
    \label{ceclam}
\end{figure}

For each value of the inflaton field $\phi$ (given by $y_1 = \sqrt{3} \tanh(\phi/\sqrt{6})$), the $\beta$-functions are given by Eq.~\eqref{betafunc}, which we use to obtain the running couplings $\lambda_1(\mu)$, $\lambda_2(\mu)$: we choose $\mu=M_0=1.3\times 10^{-5}$. The couplings at the initial scale $\mu_0=\Lambda$ are $\lambda(\mu_0)=1$, $M(\mu)=M_0$. We will take $\Lambda=1$. In Fig.~\ref{ceclam} we show the deviations $\tilde{\lambda}_i\equiv \log_{10}|\lambda_i-1| $,  in terms of $\phi$. We see that the  deviations are small, and they do not diverge for large field values.

\begin{figure}[ht]
\centering\includegraphics[width=.65\textwidth]{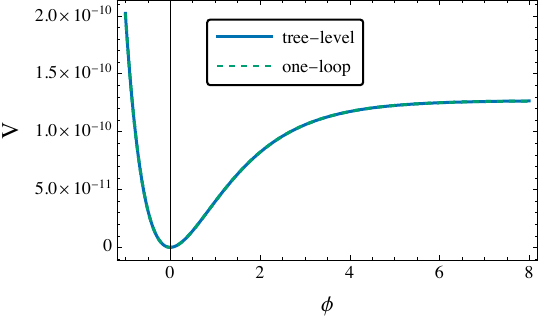}
    \caption{Comparison of the tree-level scalar potential and its one-loop corrected form in the Cecotti model~\cite{Cecotti}.}
    \label{cecv}
\end{figure}

 As one could expect from the small running of the parameters, the one-loop corrected scalar potential is essentially indistinguishable from the tree-level one, as seen in Fig.~\ref{cecv}. Indeed, the one-loop correction is always of order $M^4$. We note also that the gravitino mass vanishes, because we set $y_2=0$, and the supersymmetry-breaking order parameter $F$ stays almost constant for large field values, because $V=|F|^2$, which is different from the previous model. As a result the successful predictions of the Starobinsky model for $n_s \simeq 0.965$ and $r \simeq 0.0035$ remain unaffected by the radiative corrections considered here. 
 
\section{Summary and Conclusions}
\label{sec:conx}

In this paper we have extended our previous calculations of one-loop radiative corrections in inflationary models~\cite{Ellis:2025bzi} to include those based on supersymmetry, both global and local, i.e., supergravity. In the case of global supersymmetry, we have shown how one-loop corrections to the effective action contain logarithmic singularities that correspond to the wave-function renormalization of chiral superfields as well as including the CW effective potential. One-loop corrections to the effective potential in supergravity models with minimal kinetic terms have similar properties~\cite{hrr, Kawasaki:2000yn}. A general feature of these radiative corrections is that they do not alter substantially the tree-level predictions of the models for CMB observables. As such, they cannot account for the apparent difference in the value of $n_s$ between ACT \cite{ACT:2025fju}, SPT \cite{SPT-3G:2025bzu} and Planck \cite{Planck}.

We have then turned our attention to inflationary scenarios based on no-scale supergravity, focusing on models whose tree-level predictions are similar to those of the Starobinsky model. The general framework for such models was discussed in~\cite{Ellis:2018zya}. We have identified two model features that can significantly affect the importance of loop corrections to the inflaton potential. These are the magnitude of the gravitino mass and whether it grows with the inflaton field value, and the degree of singularity of the loop correction to the Kähler metric. If the gravitino mass grows with the inflaton field, or if the loop correction to the Kähler metric is more singular than the tree-level metric, we find that the loop correction to the inflaton potential becomes large for inflaton field values $\phi\gtrsim 6$, rendering unreliable tree-level model CMB predictions.
However, if the gravitino mass does not grow with the inflaton field, or if the one-loop Kähler metric has the same degree of singularity as the tree-level metric, the loop correction to the inflaton potential is negligible for all inflaton field values of interest, providing a substantial margin for successful Starobinsky-like inflation. 

We have derived conditions on the superpotential for the class of Starobinsky-like inflationary models in which radiative corrections remain small, namely those with small gravitino masses and/or relatively mild singularities in the Kähler metric. Among the models that have attracted attention in the literature, the Cecotti model~\cite{Cecotti} satisfies both conditions for small radiative corrections, and we also identify a class of Cecotti-like generalizations that share these desirable properties. By contrast, the original Starobinsky-like no-scale supergravity model with a Wess-Zumino superpotential~\cite{eno6} does not belong to this class.

The radiative corrections considered in this paper are all derived from the inflaton superfield self-interactions in the framework of supersymmetry and supergravity. However, the inflaton must couple to other fields in order to reheat the Universe after inflation,  the simplest mechanism being direct inflaton decays to Standard Model fields. Radiative corrections due to the coupling of the inflaton to matter fermions and scalars lead to corrections to the potential that modify the value of $n_s$. Consistency with CMB data leads to constraints on the inflaton couplings and the masses of the fields coupled to the inflaton~\cite{Ellis:2025bzi} that depend on the measurements considered~\cite{Planck,ACT:2025fju,SPT-3G:2025bzu}. We leave for future work a discussion of these constraints in the context of supergravity.~\footnote{See \cite{Ema:2024sit} for a discussion of reheating and inflaton decays in the context of no-scale supergravity.}

It has long been emphasized that supersymmetry has features that are desirable for successful inflation, through the control of radiative corrections that it provides~\cite{ENOT}. Our analysis supports this argument, and has provided a criterion for identifying Starobinsky-like no-scale supergravity models where the one-loop radiative corrections are completely negligible.

\section*{Acknowledgments}

The work of J.E. was supported by the United Kingdom STFC Grant ST/T000759/1.
The work of T.G. and K.A.O. was supported in part by DOE grant DE-SC0011842 at the University of Minnesota. The work of K.K. was supported in part by Niigata University Grant for the Enhancement of International Collaborative Research, 2025.

\clearpage

\appendix

\section{One-loop corrections to the Kähler potential in specific models}
\label{app:k1}

In this appendix we provide complete expressions for the one-loop corrections to the Kähler potentials in the KYY model as well as in the no-scale Wess-Zumino and Cecotti models, using the following abbreviated notations

\begin{equation}\begin{aligned}
    &
    \mathcal{M}\equiv \frac{M^2}{32\pi^2}e^{K^{(0)}}\log\mu^2 ,\quad Y_1\equiv  |y_ 1|^2-3,\quad Y_2\equiv  |y_ 2|^2-3, \\&\mathcal{Y}_{\al,\beta,\gamma}\equiv \alpha\Lambda^2+\beta|y_2|^2+\gamma|y_2|^4 ,\quad\Phi_i=\Phi-\bar{\Phi} \, .
\end{aligned}
\label{KYYK1}
\end{equation}

\noindent \textbf{KYY model:}
\begin{equation}
    \begin{aligned}
        K^{(1)}=&\frac{\mathcal{M}}{9}\left\{e^{-\sqrt{\frac{2}{3}}\lambda(\Phi+\bar{\Phi})}\left[9 |S|^6 +6 \left
 ( |S|^4 + 1 \right) \left(2 \lambda^2 - 3   \Phi_i^2+3\right) \right.  \right.\\&\left.\left.+ 
|S|^2(4 \lambda^4 + 
    9 ( \Phi_i^4- 4 \Phi_i^2-3  ) - 
    12 \lambda^2 \Phi_i^2)
    \right]  \right.\\&\left.+
 9   \left[|S|^6 - 2 |S|^4 ( \Phi_i^2-1)  + 
 |S|^2(  \Phi_i^4- 
       4 \Phi_i^2-3 )- 
    2 (  \Phi_i^2+1)\right]   \right\} \\& + \frac{\mathcal{M}}{9} \left\{
 3 e^{-\sqrt {\frac {2} {3}} \lambda \Phi} \left[6 - 
    3 |S|^6 + 
  \left(  6 |S|^4- 
    2\lambda^2 |S|^2
  \right)( \Phi_i^2-1) \right.\right.\\&\left.\left.  - 
    3|S|^2(  \Phi_i^4- 
       4 \Phi_i^2 -3)- 
    2 \sqrt {6} \lambda (-1 + |S|^2(-2 - 
         |S|^2+ \Phi_i^2)) \Phi_i
+ 6 \Phi_i^2\right] +\text{h.c.}
        \right\}\,.
        \label{KYYK1}
    \end{aligned}
\end{equation}

\noindent \textbf{No-scale Wess-Zumino model:}
\begin{equation}
    \begin{aligned}
        K^{(1)}&=\frac{\mathcal{M}^2e^{4K^{(0)}/3}}{18(3\Lambda^2)^9}\left\{   -54\Lambda ^{10}\mathcal{Y}_{Y_1,\Lambda^2,-3} ^4
        \left|\mathcal{K}_1\right|^2 +    \frac{3\Lambda^6}{\mathcal{Y}_{1,-12,1}^2}\mathcal{Y}_{Y_1,-12Y_1,-3}   \mathcal{Y}_{-3,\Lambda^2,-3}       \mathcal{Y}_{Y_1,\Lambda^2,-3}    ^2\  \left|\mathcal{K}_2\right|^2\right.\\&\left. +\frac{ \Lambda^6}{3\mathcal{Y}_{1,-12,1}^2} \mathcal{Y}_{Y_1,\Lambda^2,-3}^2\left(2\mathcal{Y}_{Y_1,-12Y_1,-3}^2 \left|\mathcal{K}_3\right|^2 +2  \mathcal{Y}_{-3,\Lambda^2,-3}    ^2\left|\mathcal{K}_4\right|^2+ \mathcal{Y}_{1,-6,0}^2|y_1y_2|^2\left|\mathcal{K}_5\right|^2\right) \right. \\&  \left. -\frac{27 \Lambda^8\mathcal{Y}_{Y_1,\Lambda^2,-3}^4}{ \mathcal{Y}_{1,-12,1}}\mathcal{K}_6 +\frac{2\Lambda ^6 \mathcal {Y} _ {1, -6, 
    0}^2 \mathcal {Y} _ {Y_ 1, \Lambda^2, -3}^2 }{ \mathcal{Y}_{1,-12,1}^2}\left(y_1^2\ytb^2\mathcal{K}_7+\text{h.c.} \right)\right. \\&  \left.+\frac{2\Lambda ^6 \mathcal {Y} _ {1, -6, 
    0} \mathcal {Y} _ {Y_ 1, \Lambda^2, -3}^2 \mathcal {Y} _ {-3, \Lambda^2, -3}}{ \mathcal{Y}_{1,-12,1}^2}\left(y_1 \ytb \mathcal{K}_8+\text{h.c.} \right)  \right. \\&  \left.+\frac{2\Lambda^6 \mathcal {Y} _ {1, -6, 
    0} \mathcal {Y} _ {Y_ 1, \Lambda^2, -3}^2 \mathcal {Y} _ {Y_1, -12Y_1, -3}}{ \mathcal{Y}_{1,-12,1}^2}\left(y_1 \ytb \mathcal{K}_9+\text{h.c.} \right)\right\}\,,
    \label{WZK1}
    \end{aligned}
\end{equation}
with
\begin{equation}
    \begin{aligned}
        \mathcal{K}_1&=18 c^\prime y_2^2 + y_1^2 \left[9 \lambda_1 + \sqrt{3} \left(  3 \lambda_3y_2-2 \lambda_2 y_1  \right)\right]  \end{aligned}
\end{equation}\begin{equation}
    \begin{aligned}        \mathcal{K}_2&=-2 \sqrt {3} \lambda_ 3\yob \mathcal{Y}_{Y_1,\Lambda^2,-3}^2    + 
 y_ 1 \Lambda^2 \left (36 c^\prime \ytb + 
    3 \sqrt {3} \yob^2 \lambda_ 3\right) \mathcal{Y}_{Y_1,\Lambda^2,-3} \\&+ 
 6 \yob y_ 2 \mathcal{Y}_{ 1,-6,0}\mathcal{Y}_{Y_1,\Lambda^2,-3}  \left[3 \lambda_ 1 + \sqrt {3} \left(  
 \ytb \lambda_ 3-\yob \lambda_ 2\right)\right] \\&- 
 4 y_ 1 y_ 2 \Lambda^2  \mathcal{Y}_{ 1,-6,0} \left[18 c^\prime \ytb^2 + \yob^2 \left(9 \lambda_ 1 \
+ \sqrt {3} ( 3 \ytb \lambda_ 3-2 \yob \lambda_ 2 )\right)\right] \\&+ \frac{3 y_ 1\mathcal{Y}_{ Y_1,\Lambda^2,-3}  }{\left( | y_ 1 | ^2 + | 
      y_ 2 | ^2-3\right)\left[\left( | y_ 1 | ^2 + | 
      y_ 2 | ^2-3\right)Y_2\Lambda+9\right]} \left[  - \mathcal{Y}_{ Y_1,\Lambda^2,-3}  \left(36 c^\prime \ytb + 
           3 \sqrt {3} \yob^2 \lambda_ 3\right)\right. \\&  \left.+54 c^\prime \ytb | y_ 2 |^2 \mathcal{Y}_{ 1,-6,0}     + 
      3  \yob^2y_ 2 \mathcal{Y}_{ 1,-6,0}  \left(9 \
\lambda_ 1 + \sqrt {3} (  
               3 \ytb \lambda_ 3-2 \yob \lambda_ 2)\right) \right]\\&+ \frac{\mathcal{Y}_{Y_1,\Lambda^2,-3}}{\left( | y_ 1 | ^2 + | 
      y_ 2 | ^2-3\right)\left[\left( | y_ 1 | ^2 + | 
      y_ 2 | ^2-3\right)Y_2\Lambda+9\right]}\left[y_ 2 (9 + (Y_ 2^2 - | 
         y_ 1 | ^4) \Lambda)  \right. \\&\left.(18 c^\prime y_ 1 \ytb^2 \Lambda^2 +  
\yob (6 | 
         y_ 2 | ^4 (3 \lambda_ 1 + \sqrt {3} (  
\ytb \lambda_ 3-\yob \lambda_ 2))\right. \\&\left.- 2 | 
         y_ 2 | ^2 \Lambda^2 (3 \lambda_ 1 + \sqrt {3} (    \ytb \lambda_ 3- \yob  
\lambda_ 2)) + \Lambda^2 (3 (6 + | y_ 1 | ^2) \lambda_1 + \sqrt{3} (-6 \yob \lambda_2 + (6 + | y_ 1 | ^2) \ytb \lambda_ 3))))\right]  
 \\      \mathcal{K}_3&=-3 \mathcal {Y} _ {Y_ 1, \Lambda^2, -3}^2 \left[ 3 \lambda_ 1 + 
     \sqrt{3} (  y_ 2 \lambda_ 3-2 y_ 1 \lambda_ 2)\right] \\&+ 18 | 
 y_ 1 | ^2 \Lambda^2  \mathcal {Y} _ {Y_ 1, \Lambda^2, -3}  \left[3 \lambda_ 1 + \sqrt{3} (y_ 2 \lambda_ 3 - y_ 1 \lambda_ 2 )\right] \\&- 
  6 \yob^2 \Lambda^4  \left[18 c^\prime y_ 2^2 + 
     y_ 1^2 (9 \lambda_ 1 + 
      \sqrt{3}  ( 
           3 y_ 2 \lambda_ 3-2 y_ 1 \lambda_ 2 ))\right] \\&+ \frac{1}{| 
     y_ 1 | ^2 + | y_ 2 | ^2 - 3 }\left[\yob  \mathcal {Y} _ {Y_ 1,  
\Lambda^2, -3}  (-6 y_ 1 \mathcal {Y} _ {Y_ 1, \Lambda^2, -3}  (3  
\lambda_ 1 + \sqrt{3} (y_ 2 \lambda_3 - y_ 1 \lambda_ 2 )) \right.\\&\left.+ 
       3 \yob \Lambda^2 (18 c^\prime y_ 2^2 + 
          y_ 1^2 (9 \lambda_ 1 + 
            \sqrt{3} (3 y_ 2 \lambda_ 3 - 2 y_ 1 \lambda_ 2))))\right]  \\        \mathcal{K}_4&=-18 c^\prime \mathcal{Y}_{Y_1,\Lambda^2,-3}^2+9 \ytb \mathcal{Y} _{1, -6, 
   0} \mathcal{Y}_ {Y_ 1, \Lambda^2, -3} \left(12 c^\prime y_2 + \sqrt{3} \lambda_ 3y_1^2 \right)\\&-  \ytb^2 \left(6\mathcal{Y} _ {1, -6, 0}^2+9\mathcal {Y} _ {Y_ 1, \Lambda^2, -3}  \right)\left[18 c^\prime y_ 2^2 + 
   y_ 1^2 (9 \lambda_ 1 + \sqrt{3} (3 y_ 2 \lambda_ 3 - 
         2 y_ 1 \lambda_ 2))\right]\\&-\frac{y_ 1 \ytb^2 \Lambda \mathcal {Y} _ {Y_ 1, \Lambda^2, -3}}{ \left( | y_ 1 | ^2 + | 
      y_ 2 | ^2-3\right)Y_2\Lambda+9}\left[
      -6 y_ 1  \mathcal {Y} _ {Y_ 1, \Lambda^2, -3} (3 \lambda_ 1 + \sqrt{3} (y_ 2 \lambda_ 3 - y_ 1 \lambda_ 2)) \right.\\&\left.+ 
 3 \yob \Lambda^2 (18 c^\prime y_ 2^2 + 
    y_ 1^2 (9 \lambda_ 1 + \sqrt {3} (3 y_ 2 \lambda_ 3 - 
          2 y_ 1 \lambda_ 2)))\right]
\\&+ \frac{9 \ytb \mathcal {Y} _ {Y_ 1, \Lambda^2, -3}}{\left( | y_ 1 | ^2 + | 
      y_ 2 | ^2-3\right)\left[\left( | y_ 1 | ^2 + | 
      y_ 2 | ^2-3\right)Y_2\Lambda+9\right]}      \left[3 \ytb \mathcal {Y} _ {1, -6, 
   0} (18 c^\prime y_ 2^2 \right.\\&\left.+ 
    y_ 1^2 (9 \lambda_ 1 + \sqrt {3} (3 y_ 2 \lambda_ 3 - 
          2 y_ 1 \lambda_ 2))) - \mathcal {Y} _ {Y_ 1, \Lambda^2, -3} \
(36 c^\prime y_ 2 + 3 \sqrt {3} y_ 1^2 \lambda_ 3)\right]   \\        \mathcal{K}_5&=-6 \sqrt {3} y_ 1\lambda_ 3 \mathcal {Y} _ {Y_ 1, \Lambda^2, -3}^2+3 \yob \Lambda^2 \mathcal {Y} _ {Y_ 1, \Lambda^2, -3} \left(36 \
c^\prime y_ 2 + 3 \sqrt {3} y_ 1^2 \lambda_ 3\right)
\\&+18 y_ 1 \ytb \mathcal {Y} _ {1, -6,  
0}  \mathcal {Y} _ {Y_ 1, \Lambda^2, \
-3} \left[3 \lambda_ 1 + \sqrt {3} (  y_ 2 \lambda_ 3- y_ 1 \lambda_ 2 )\right]\\&-12 \yob \ytb \Lambda^2\mathcal {Y} _ {1, -6,  
0}   \left[18 c^\prime y_ 2^2 + 
   y_ 1^2 (9 \lambda_ 1 + \sqrt {3} (  
         3 y_ 2 \lambda_ 3-2 y_ 1 \lambda_ 2))\right]\\&+ \frac{ \ytb \mathcal {Y} _ {Y_ 1, \Lambda^2, -3}  \left[9 + (Y_2^2 - |y_ 1|^4) 
      \Lambda\right]  
      }{\left( | y_ 1 | ^2 + | 
      y_ 2 | ^2-3\right)\left[\left( | y_ 1 | ^2 + | 
      y_ 2 | ^2-3\right)Y_2\Lambda+9\right]}\left[-6 y_ 1\mathcal {Y} _ {Y_ 1, \Lambda^2, -3}  (3 \lambda_ 1 + \sqrt {3} (  y_ 2 \lambda_ 3- y_ 1 \lambda_ 
2 ))\right.\\&\left. + 
  3 \yob \Lambda^2 (18 c^\prime y_ 2^2 + 
     y_ 1^2 (9 \lambda_ 1 + \sqrt {3} (-2 y_ 1 \lambda_ 2 + 
           3 y_ 2 \lambda_ 3)))\right]\\&+ \frac{ 9 \yob \mathcal {Y} _ {Y_ 1, \Lambda^2, -3} }{\left( | y_ 1 | ^2 + | 
      y_ 2 | ^2-3\right)\left[\left( | y_ 1 | ^2 + | 
      y_ 2 | ^2-3\right)Y_2\Lambda+9\right]}\times\\&\left[3 \ytb  \mathcal {Y} _ {1, -6, 
     0} (18 c^\prime y_ 2^2 + 
      y_ 1^2 (9 \lambda_ 1 + \sqrt {3} ( 
            3 y_ 2 \lambda_ 3-2 y_ 1 \lambda_ 2 ))) - \mathcal {Y} _ {Y_ 1, \Lambda^2, \
-3} (36 c^\prime y_ 2 + 3 \sqrt {3} y_ 1^2 \lambda_ 3) \right] \end{aligned}
\end{equation}\begin{equation}
    \begin{aligned}          \mathcal{K}_6&=\left[2 c^\prime \ytb \mathcal{Y}_{6,\Lambda^2,-12}  + \yob^2 (y_ 2 \Lambda^2 \lambda_ 1 + \
\sqrt {3} \Lambda^2 \lambda_ 3 - 
      y_ 2^2 \ytb (6 \lambda_ 1 + \sqrt {3} \ytb\lambda_ 3))\right]\times \\&\left[-  \mathcal{Y}_{Y_1,\Lambda^2,-3}   (36 c^\prime y_ 2 + 
        3 \sqrt {3} y_ 1^2 \lambda_ 3)  + 
   3 \ytb \mathcal{Y}_{1,-6,0}   (18 c^\prime y_ 2^2 + 
      y_ 1^2 (9 \lambda_ 1 + \sqrt {3} ( 
            3 y_ 2 \lambda_ 3-2 y_ 1 \lambda_ 2 )))\right]\\&+\left[-6 y_ 1 \mathcal{Y}_{Y_1,\Lambda^2,-3}    (3 \lambda_ 1 + \sqrt {3} (  y_ 2 \lambda_ 3-y_ 1  
\lambda_ 2 )) + 
     3 \yob\Lambda^2 (18 c^\prime y_ 2^2 + 
        y_ 1^2 (9 \lambda_ 1 + \sqrt {3} ( 
              3 y_ 2 \lambda_ 3-2 y_ 1 \lambda_ 2)))\right] \\&
              \times\left[2 c^\prime y_ 1 \ytb^2 \mathcal{Y}_{1,-24,0} + \yob (2 | 
       y_ 2 | ^4 (3 \lambda_ 1 + \sqrt {3} (  \
\ytb\lambda_ 3-\yob\lambda_ 2))\right.\\&\left. + \Lambda^2 ((6 + | y_ 1 | ^2) \lambda_ 1 + 
           2 \sqrt {3} (  \ytb\lambda_ 3-\yob\lambda_ 2)) - 2 | 
       y_ 2 | ^2 (6 (6 + | 
              y_ 1 | ^2) \lambda_ 1 \right.\\&\left.+ \sqrt {3} (-12 \yob\lambda_ 2 + \
(12 + | y_ 1 | ^2) \ytb\lambda_ 3)))\right]\\        \mathcal{K}_7&= \left[-3 \mathcal {Y} _ {Y_ 1, \Lambda^2, -3}^2 (3 \lambda_ 1 + \sqrt{3} ( y_ 2 \lambda_ 3-2 y_ 1 \lambda_ 2 )) + 18 | 
   y_ 1 | ^2 \Lambda^2 \mathcal {Y} _ {Y_ 1, \Lambda^2, -3} (3 
\lambda_ 1 + \sqrt {3} (y_ 2 \lambda_ 3 - y_ 1 \lambda_ 2))  \right. \\& \left.  - 
    6 \yob^2 \Lambda^4 (18 c^\prime y_ 2^2 + 
       y_ 1^2 (9 \lambda_ 1 + \sqrt{3} (3 y_ 2 \lambda_ 3 - 
             2 y_ 1 \lambda_ 2))) \right. \\& \left. + (\yob \mathcal {Y} _ {Y_ 1,  
\Lambda^2, -3} (-6 y_ 1 \mathcal{Y}_{Y_1,\Lambda^2,-3} (3 \lambda_ 1 + \sqrt {3} (y_ 2 \
\lambda_ 3 - y_ 1 \lambda_ 2))  \right. \\& \left.  + 
         3 \yob \Lambda^2 (18 c^\prime y_ 2^2 + 
            y_ 1^2 (9 \lambda_ 1 + \sqrt {3} (3 y_ 2 \lambda_ 3 - 
                  2 y_ 1 \lambda_ 2)))))   /(-3 + | y_ 1 | ^2 + | 
       y_ 2 | ^2)\right]\times  \\&  \left[-6 c^\prime \mathcal {Y} _ {Y_ 1, \Lambda^2, 
-3}^2 + 3 y_ 2 \mathcal {Y} _ {1, -6, 
      0} \mathcal {Y} _ {Y_ 1, \Lambda^2, -3} (12 c^\prime \ytb + \
\sqrt {3} \yob^2 \lambda_ 3) \right. \\& \left. - 
   2 y_ 2^2 \mathcal {Y} _ {1, -6, 
      0}^2 (18 c^\prime \ytb^2 + \yob^2 (9 \lambda_ 1 + \sqrt {3} (3 
\ytb \lambda_ 3 - 2 \yob \lambda_ 2))) \right. \\& \left. - 
   3 y_ 2^2 \mathcal {Y} _ {Y_ 1, \Lambda^2, -3} (18 c^\prime \ytb^2 \
+ \yob^2 (9 \lambda_ 1 + \sqrt {3} (3 \ytb \lambda_ 3 - 
            2 \yob \lambda_ 2))) \right. \\& \left. + (3 y_ 2 \mathcal {Y} _ {Y_ 1, \
\Lambda^2, -3} (-(\mathcal {Y} _ {Y_ 1, \Lambda^2, -3} (36 c^\prime \
\ytb + 3 \sqrt{3} \yob^2 \lambda_ 3)) \right. \\& \left. + 
        3 y_ 2 \mathcal {Y} _ {1, -6, 
          0} (18 c^\prime \ytb^2 + \yob^2 (9 \lambda_ 1 + \sqrt{3} (3 \ytb \lambda_ 3 - 2 \yob \lambda_ 2))))) \frac{1}{\left( | y_ 1 | ^2 + | 
      y_ 2 | ^2-3\right)\left[\left( | y_ 1 | ^2 + | 
      y_ 2 | ^2-3\right)Y_2\Lambda+9\right]} \right. \\& \left.  - (\yob y_ 2^2 \Lambda \mathcal {Y} 
_ {Y_ 1, \Lambda^2, -3} (18 c^\prime y_ 1 \ytb^2 \Lambda^2 + \yob (6 |
            y_ 2 | ^4 (3 \lambda_ 1 + \sqrt {3} (  
\ytb \lambda_ 3- \yob \lambda_ 2 )) \right. \\& \left.  - 2 | 
           y_ 2 | ^2 \Lambda^2 (3 \lambda_ 1 + \sqrt {3} (  \ytb \lambda_ 3- \yob 
\lambda_ 2 )) + \Lambda^2 (3 (6 + | 
                   y_ 1 | ^2) \lambda_ 1  \right. \\& \left. + \sqrt{3} (-6 \yob 
\lambda_ 2 + (6 + | y_ 1 | ^2) \ytb \lambda_ 3)))))/(9 + Y_ 2 (-3 + | y_ 1 | ^2 + | y_ 2 | ^2) \Lambda)\right]
\end{aligned}
\end{equation}\begin{equation}
    \begin{aligned}           \mathcal{K}_8&=\left[-6 \sqrt{3} y_ 1 \mathcal{Y}_{Y_ 1,\Lambda^2,-3}^2 \lambda_3 + 3 \yob \Lambda^2 \mathcal{Y}_{Y_ 1,\Lambda^2,-3} (36c^\prime y_ 2 + 3 \sqrt{3} y_ 1^2 \lambda_3) \right. \\& \left. + 18 y_ 1 \ytb \mathcal{Y}_{1,-6,0} \mathcal{Y}_{Y_ 1,\Lambda^2,-3} (3 \lambda_1 + \sqrt{3} (  y_ 2 \lambda_3-y_ 1 \lambda_2 )) \right. \\& \left. - 12 \yob \ytb \Lambda^2 \mathcal{Y}_{1,-6,0} (18c^\prime y_ 2^2 + y_ 1^2 (9 \lambda_1 + \sqrt{3} (  3 y_ 2 \lambda_3-2 y_ 1 \lambda_2))) \right. \\& \left. + (\ytb (9 + (Y_ 2^2 - |y_ 1|^4) \Lambda) \mathcal{Y}_{Y_ 1,\Lambda^2,-3} (6 y_ 1 (3 |y_ 2|^4 - Y_ 1 \Lambda^2 - |y_ 2|^2 \Lambda^2) (3 \lambda_1 + \sqrt{3} (-(y_ 1 \lambda_2) + y_ 2 \lambda_3)) \right. \\& \left. + 3 \yob \Lambda^2 (18c^\prime y_ 2^2 + y_ 1^2 (9 \lambda_1 + \sqrt{3} (  3 y_ 2 \lambda_3-2 y_ 1 \lambda_2)))))\frac{1}{\left( | y_ 1 | ^2 + | 
      y_ 2 | ^2-3\right)\left[\left( | y_ 1 | ^2 + | 
      y_ 2 | ^2-3\right)Y_2\Lambda+9\right]}  \right. \\& \left.  + (9 \yob \mathcal{Y}_{Y_ 1,\Lambda^2,-3} (-(\mathcal{Y}_{Y_ 1,\Lambda^2,-3} (36c^\prime y_ 2 + 3 \sqrt{3} y_ 1^2 \lambda_3))  \right. \\& \left. + 3 \ytb \mathcal{Y}_{1,-6,0} (18c^\prime y_ 2^2 + y_ 1^2 (9 \lambda_1 + \sqrt{3} ( 3 y_ 2 \lambda_3-2 y_ 1 \lambda_2 )))))\frac{1}{\left( | y_ 1 | ^2 + | 
      y_ 2 | ^2-3\right)\left[\left( | y_ 1 | ^2 + | 
      y_ 2 | ^2-3\right)Y_2\Lambda+9\right]}  \right]\\ &\times    
      \left[-6c^\prime \mathcal{Y}_{Y_ 1,\Lambda^2,-3}^2 + 3 y_ 2 \mathcal{Y}_{1,-6,0} \mathcal{Y}_{Y_ 1,\Lambda^2,-3} (12c^\prime \ytb + \sqrt{3} \yob^2 \lambda_3)   \right. \\& \left. - 2 y_ 2^2 \mathcal{Y}_{1,-6,0}^2 (18c^\prime \ytb^2 + \yob^2 (9 \lambda_1 + \sqrt{3} (-2 \yob \lambda_2 + 3 \ytb \lambda_3)))   \right. \\& \left. - 3 y_ 2^2 \mathcal{Y}_{Y_ 1,\Lambda^2,-3} (18c^\prime \ytb^2 + \yob^2 (9 \lambda_1 + \sqrt{3} (-2 \yob \lambda_2 + 3 \ytb \lambda_3)))   \right. \\& \left. + (3 y_ 2 \mathcal{Y}_{Y_ 1,\Lambda^2,-3} (-(\mathcal{Y}_{Y_ 1,\Lambda^2,-3} (36c^\prime \ytb + 3 \sqrt{3} \yob^2 \lambda_3))  \right. \\& \left.  + 3 y_ 2 \mathcal{Y}_{1,-6,0} (18c^\prime \ytb^2 + \yob^2 (9 \lambda_1 + \sqrt{3} (-2 \yob \lambda_2 + 3 \ytb \lambda_3)))))\frac{1}{\left( | y_ 1 | ^2 + | 
      y_ 2 | ^2-3\right)\left[\left( | y_ 1 | ^2 + | 
      y_ 2 | ^2-3\right)Y_2\Lambda+9\right]} \right. \\& \left. - (\yob y_ 2^2 \Lambda \mathcal{Y}_{Y_ 1,\Lambda^2,-3} (18c^\prime y_ 1 \ytb^2 \Lambda^2 + \yob (6 |y_ 2|^4 (3 \lambda_1 + \sqrt{3} ( \ytb \lambda_3-\yob \lambda_2)) \right. \\& \left. - 2 |y_ 2|^2 \Lambda^2 (3 \lambda_1 + \sqrt{3} (  \ytb \lambda_3- \yob \lambda_2  )) + \Lambda^2 (3 (6 + |y_ 1|^2) \lambda_1 \right. \\& \left. + \sqrt{3} (-6 \yob \lambda_2 + (6 + |y_ 1|^2) \ytb \lambda_3)))))/(9 + Y_ 2 (-3 + |y_ 1|^2 + |y_ 2|^2) \Lambda)\right]    \\\mathcal{K}_9&= \left[-3 \mathcal {Y} _ {Y_ 1, \Lambda^2, -3}^2 (3 \lambda_ 1 + \sqrt {3} (y_ 2 \lambda_ 3 - 2 y_ 1 \lambda_ 2)) + 18 | y_ 1 | ^2 \Lambda^2 \mathcal {Y} _ {Y_ 1, \Lambda^2, -3} (3 \lambda_ 1 + \sqrt {3} (y_ 2 \lambda_ 3 - y_ 1 \lambda_ 2)) \right. \\& \left. - 6 \yob^2 \Lambda^4 (18 c^\prime y_ 2^2 + y_ 1^2 (9 \lambda_ 1 + \sqrt {3} (3 y_ 2 \lambda_ 3 - 2 y_ 1 \lambda_ 2))) \right. \\& \left. + (\yob \mathcal {Y} _ {Y_ 1, \Lambda^2, -3} (-6 y_ 1 \mathcal {Y} _ {Y_ 1, \Lambda^2, -3} (3 \lambda_ 1 + \sqrt {3} (y_ 2 \lambda_ 3 - y_ 1 \lambda_ 2)) \right. \\& \left. + 3 \yob \Lambda^2 (18 c^\prime y_ 2^2 + y_ 1^2 (9 \lambda_ 1 + \sqrt {3} (3 y_ 2 \lambda_ 3 - 2 y_ 1 \lambda_ 2)))))/(-3 + | y_ 1 | ^2 + | y_ 2 | ^2)\right]\times\\&\left[-2 \sqrt {3} \yob \mathcal {Y} _ {Y_ 1, \Lambda^2, -3}^2 \lambda_ 3 + y_ 1 \Lambda^2 \mathcal {Y} _ {Y_ 1, \Lambda^2, -3} (36 c^\prime \ytb + 3 \sqrt {3} \yob^2 \lambda_ 3) \right. \\& \left.+ 6 \yob y_ 2 \mathcal {Y} _ {1, -6, 0} \mathcal {Y} _ {Y_ 1, \Lambda^2, -3} (3 \lambda_ 1 + \sqrt {3} (\ytb \lambda_ 3 - \yob \lambda_ 2))\right. \\& \left. - 4 y_ 1 y_ 2 \Lambda^2 \mathcal {Y} _ {1, -6, 0} (18 c^\prime \ytb^2 + \yob^2 (9 \lambda_ 1 + \sqrt {3} (3 \ytb \lambda_ 3 - 2 \yob \lambda_ 2)))\right. \\& \left. + (3 y_ 1 \mathcal {Y} _ {Y_ 1, \Lambda^2, -3} (-(\mathcal {Y} _ {Y_ 1, \Lambda^2, -3} (36 c^\prime \ytb + 3 \sqrt {3} \yob^2 \lambda_ 3)) \right. \\& \left.+ 3 y_ 2 \mathcal {Y} _ {1, -6, 0}(18 c^\prime \ytb^2 + \yob^2 (9 \lambda_ 1 + \sqrt {3} (3 \ytb \lambda_ 3 - 2 \yob \lambda_ 2)))))\frac{1}{\left( | y_ 1 | ^2 + | 
      y_ 2 | ^2-3\right)\left[\left( | y_ 1 | ^2 + | 
      y_ 2 | ^2-3\right)Y_2\Lambda+9\right]} \right. \\& \left.+ (y_ 2 (9 + (Y_ 2^2 - | y_ 1 | ^4) \Lambda) \mathcal {Y} _ {Y_ 1, \Lambda^2, -3} (18 c^\prime y_ 1 \ytb^2 \Lambda^2 + \yob (6 | y_ 2 | ^4 (3 \lambda_ 1 + \sqrt {3} (\ytb \lambda_ 3 - \yob \lambda_ 2)) \right. \\& \left.- 2 | y_ 2 | ^2 \Lambda^2 (3 \lambda_ 1 + \sqrt {3} (\ytb \lambda_ 3 - \yob \lambda_ 2)) + \Lambda^2 (3 (6 + | y_ 1 | ^2) \lambda_ 1 \right. \\& \left.+ \sqrt {3} (-6 \yob \lambda_ 2 + (6 + | y_ 1 | ^2) \ytb \lambda_ 3)))))\frac{1}{\left( | y_ 1 | ^2 + | 
      y_ 2 | ^2-3\right)\left[\left( | y_ 1 | ^2 + | 
      y_ 2 | ^2-3\right)Y_2\Lambda+9\right]} \right]
    \end{aligned}
\end{equation}
\newpage
\noindent \textbf{No-scale Cecotti model:}

The complete expression of $K^{(1)}$ in the Cecotti model is very lengthy, so
we instead present the one-loop correction  in the limit of a large stabilization parameter $\Lambda\rightarrow\infty$:
\begin{equation}
    \begin{aligned}
        K^{(1)}=&-\frac{2\mathcal{M} }{27}\left\{\lambda_ 1^2
       \left[|y_ 1|^6 + 
     2 |y_ 1|^4 Y_2 + Y_2 \left(9 + 
      |y_ 2|^2 Y_2\right) + 
     2 |y_ 1|^2 \left(Y_2^2+42Y_2+126\right)\right] \right.\\&\left.+ \sqrt {3}\lambda_ 1 \lambda_ 2 (y_ 1 + \bar
{y} _ 1) \left[ 
    |y_ 1|^2 ( |y_ 1|^2  - 
        26 |y_ 2|^2-9 )-6Y_2\right]  \right.\\&\left.+  \lambda_ 2^2\left[-3 |y_ 1|^6 + 
     |y_ 1|^4 (21 + 22 |y_ 2|^2)- 4 |y_ 1|^2 Y_2^2 - 
     2 |y_ 2|^2 Y_2^2 \right]\right\} \, .
    \end{aligned}
    \label{CecottiK1}
\end{equation}
We have verified that using the exact, finite $\Lambda$ expression or the infinite $\Lambda$ expression \eqref{CecottiK1}  gives similar negligible effects, so our conclusions for the Cecotti model do not depend on this choice.


\begin{thebibliography}{99}

\bibitem{Planck}
N.~Aghanim \textit{et al.} [Planck Collaboration],
Astron. Astrophys. \textbf{641}, A6 (2020)
[erratum: Astron. Astrophys. \textbf{652}, C4 (2021)]
[arXiv:1807.06209 [astro-ph.CO]];
Y.~Akrami \textit{et al.} [Planck Collaboration],
Astron. Astrophys. \textbf{641}, A10 (2020)
[arXiv:1807.06211 [astro-ph.CO]].


\bibitem{Staro}
A.~A.~Starobinsky,
Phys.\ Lett.\ B {\bf 91}, 99 (1980).

\bibitem{eno6}
J.~Ellis, D.~V.~Nanopoulos and K.~A.~Olive,
Phys. Rev. Lett. \textbf{111}, 111301 (2013)
[erratum: Phys. Rev. Lett. \textbf{111}, no.12, 129902 (2013)]
[arXiv:1305.1247 [hep-th]].

\bibitem{Kallosh:2013lkr}
R.~Kallosh and A.~Linde,
JCAP \textbf{06}, 028 (2013)
[arXiv:1306.3214 [hep-th]].

\bibitem{Ellis:2013nxa}
J.~Ellis, D.~V.~Nanopoulos and K.~A.~Olive,
JCAP \textbf{10}, 009 (2013)
[arXiv:1307.3537 [hep-th]].

\bibitem{Ellis:2018zya}
J.~Ellis, D.~V.~Nanopoulos, K.~A.~Olive and S.~Verner,
JHEP \textbf{03}, 099 (2019)
[arXiv:1812.02192 [hep-th]].

\bibitem{Antoniadis:2025pfa}
I.~Antoniadis, J.~Ellis, W.~Ke, D.~V.~Nanopoulos and K.~A.~Olive,
JCAP \textbf{08}, 090 (2025)
[arXiv:2504.12283 [hep-ph]].

\bibitem{ACT:2025fju}
T.~Louis \textit{et al.} [Atacama Cosmology Telescope Collaboration],
JCAP \textbf{11}, 062 (2025)
[arXiv:2503.14452 [astro-ph.CO]];
E.~Calabrese \textit{et al.} [Atacama Cosmology Telescope Collaboration],
JCAP \textbf{11}, 063 (2025)
[arXiv:2503.14454 [astro-ph.CO]].

\bibitem{Kallosh:2025rni}
R.~Kallosh, A.~Linde and D.~Roest,
Phys. Rev. Lett. \textbf{135}, no.16, 161001 (2025)
[arXiv:2503.21030 [hep-th]].

\bibitem{Dioguardi:2025vci}
C.~Dioguardi, A.~J.~Iovino and A.~Racioppi,
Phys. Lett. B \textbf{868}, 139664 (2025)
[arXiv:2504.02809 [gr-qc]].

\bibitem{Gialamas:2025kef}
I.~D.~Gialamas, A.~Karam, A.~Racioppi and M.~Raidal,
Phys. Rev. D \textbf{112}, no.10, 103544 (2025)
[arXiv:2504.06002 [astro-ph.CO]].


\bibitem{Drees:2025ngb}
M.~Drees and Y.~Xu,
Phys. Lett. B \textbf{867}, 139612 (2025)
[arXiv:2504.20757 [astro-ph.CO]].

\bibitem{Zharov:2025evb}
D.~S.~Zharov, O.~O.~Sobol and S.~I.~Vilchinskii,
Phys. Rev. D \textbf{112}, no.2, 023544 (2025)
[arXiv:2505.01129 [astro-ph.CO]].

\bibitem{Haque:2025uri}
M.~R.~Haque, S.~Pal and D.~Paul,
[arXiv:2505.01517 [astro-ph.CO]].

\bibitem{Gialamas:2025ofz}
I.~D.~Gialamas, T.~Katsoulas and K.~Tamvakis,
JCAP \textbf{09}, 060 (2025)
[arXiv:2505.03608 [gr-qc]].

\bibitem{Haque:2025uga}
M.~R.~Haque and D.~maity,
Phys. Lett. B \textbf{873}, 140187 (2026)
[arXiv:2505.18267 [astro-ph.CO]].


\bibitem{Wolf:2025ecy}
W.~J.~Wolf,
[arXiv:2506.12436 [astro-ph.CO]].

\bibitem{Ahmed:2025rrg}
W.~Ahmed and M.~U.~Rehman,
Phys. Rev. D \textbf{112}, no.6, 063519 (2025)
[arXiv:2506.18077 [astro-ph.CO]].

\bibitem{Han:2025cwk}
J.~Han, H.~M.~Lee and J.~H.~Song,
[arXiv:2506.21189 [hep-ph]].

\bibitem{Pallis:2025gii}
C.~Pallis,
JCAP \textbf{09}, 061 (2025)
[arXiv:2507.02219 [hep-ph]].

\bibitem{German:2025ide}
G.~German and J.~C.~Hidalgo,
Int. J. Mod. Phys. D \textbf{35}, no.03, 2550098 (2026)
[arXiv:2508.01017 [astro-ph.CO]].


\bibitem{deform}
J.~Ellis, M.~A.~G.~Garc{\'\i}a, N.~Nagata, D.~V.~Nanopoulos and K.~A.~Olive,
JCAP \textbf{12}, 038 (2025)
[arXiv:2508.13279 [hep-ph]].

\bibitem{Aoki:2025ywt}
S.~Aoki, H.~Otsuka and R.~Yanagita,
JCAP \textbf{11}, 088 (2025)
[arXiv:2509.06739 [hep-ph]].

\bibitem{Ellis:2025bzi}
J.~Ellis, T.~Gherghetta, K.~Kaneta, W.~Ke and K.~A.~Olive,
Phys. Rev. D \textbf{112}, no.12, 123530 (2025)
[arXiv:2510.15137 [hep-ph]].


\bibitem{Ellis:2025zrf}
J.~Ellis, M.~A.~G.~Garcia, K.~A.~Olive and S.~Verner,
[arXiv:2510.18656 [hep-ph]].

\bibitem{Alexandre:2025ixz}
J.~Alexandre, L.~Heurtier and S.~Pla,
[arXiv:2511.05296 [hep-th]].

\bibitem{Iacconi:2025odq}
L.~Iacconi, S.~Bhattacharya, M.~Fasiello and D.~Wands,
[arXiv:2511.14673 [astro-ph.CO]].

\bibitem{Kallosh:2025sji}
R.~Kallosh and A.~Linde,
[arXiv:2512.02969 [hep-th]].

\bibitem{ENOT}
J.~R.~Ellis, D.~V.~Nanopoulos, K.~A.~Olive and K.~Tamvakis,
Phys. Lett. B \textbf{118} (1982), 335;
Nucl. Phys. B \textbf{221} (1983), 524-548;
Phys. Lett. B \textbf{120} (1983), 331-334.%

\bibitem{Nakayama:2011ri}
K.~Nakayama and F.~Takahashi,
JCAP \textbf{10} (2011), 033
[arXiv:1108.0070 [hep-ph]].

\bibitem{Nanopoulos:1982bv}
D.~V.~Nanopoulos, K.~A.~Olive, M.~Srednicki and K.~Tamvakis,
Phys. Lett. B \textbf{123}, 41-44 (1983).

 \bibitem{hrr}
   R.~Holman, P.~Ramond and G.~G.~Ross,
  Phys.\ Lett.\ B {\bf 137}, 343 (1984).

  \bibitem{no-scale1}
E.~Cremmer, S.~Ferrara, C.~Kounnas and D.~V.~Nanopoulos,
  Phys.\ Lett.\ B {\bf 133}, 61 (1983).

\bibitem{EKN}
J.~R.~Ellis, C.~Kounnas and D.~V.~Nanopoulos,
Nucl. Phys. B \textbf{247}, 373-395 (1984).

\bibitem{ELNT}
J.~R.~Ellis, A.~B.~Lahanas, D.~V.~Nanopoulos and K.~Tamvakis,
Phys. Lett. B \textbf{134}, 429 (1984)

\bibitem{no-scale2}
  A.~B.~Lahanas and D.~V.~Nanopoulos,
  Phys.\ Rept.\  {\bf 145} (1987) 1.

\bibitem{Witten}
E.~Witten,
  Phys.\ Lett.\ B {\bf 155} (1985) 151;
  see also
  S.~B.~Giddings, S.~Kachru and J.~Polchinski,
  Phys.\ Rev.\ D {\bf 66}, 106006 (2002)
  [hep-th/0105097];
  V.~Balasubramanian, P.~Berglund, J.~P.~Conlon and F.~Quevedo,
  JHEP {\bf 0503}, 007 (2005)
  [hep-th/0502058].

\bibitem{Kawasaki:2000yn}
M.~Kawasaki, M.~Yamaguchi and T.~Yanagida,
Phys. Rev. Lett. \textbf{85}, 3572-3575 (2000)
[arXiv:hep-ph/0004243 [hep-ph]].

\bibitem{Linde:1981mu}
A.~D.~Linde,
Phys. Lett. B \textbf{108}, 389-393 (1982)
doi:10.1016/0370-2693(82)91219-9

\bibitem{Linde:1983gd}
A.~D.~Linde,
Phys. Lett. B \textbf{129}, 177-181 (1983).


\bibitem{Kallosh:2010ug}
R.~Kallosh and A.~Linde,
JCAP \textbf{11}, 011 (2010)
[arXiv:1008.3375 [hep-th]].

\bibitem{Kallosh:2010xz}
R.~Kallosh, A.~Linde and T.~Rube,
Phys. Rev. D \textbf{83}, 043507 (2011)
[arXiv:1011.5945 [hep-th]].

\bibitem{Kallosh:2011qk}
R.~Kallosh, A.~Linde, K.~A.~Olive and T.~Rube,
Phys. Rev. D \textbf{84}, 083519 (2011)
[arXiv:1106.6025 [hep-th]].




\bibitem{Cecotti}
  S.~Cecotti,
  Phys.\ Lett.\ B {\bf 190} (1987) 86.

\bibitem{building}
J.~Ellis, M.~A.~G.~Garcia, N.~Nagata, D.~V.~Nanopoulos, K.~A.~Olive and S.~Verner,
Int. J. Mod. Phys. D \textbf{29}, no.16, 2030011 (2020)
[arXiv:2009.01709 [hep-ph]].

\bibitem{Gaillard:1993es}
M.~K.~Gaillard and V.~Jain,
Phys. Rev. D \textbf{49}, 1951-1965 (1994)
[arXiv:hep-th/9308090 [hep-th]].

\bibitem{Gaillard:1996hs}
M.~K.~Gaillard, V.~Jain and K.~Saririan,
Phys. Lett. B \textbf{387}, 520-528 (1996)
[arXiv:hep-th/9606135 [hep-th]].




\bibitem{Grisaru:1996ve}
M.~T.~Grisaru, M.~Rocek and R.~von Unge,
Phys. Lett. B \textbf{383}, 415-421 (1996)
[arXiv:hep-th/9605149 [hep-th]].

 

\bibitem{Buchbinder:1998twe}
I.~L.~Buchbinder and S.~M.~Kuzenko,
Taylor {\&} Francis, 1998,
ISBN 978-1-4200-5051-6, 978-0-7503-0506-8, 978-0-367-80253-0

\bibitem{Martin:2024qmi}
S.~P.~Martin,
Phys. Rev. D \textbf{110}, no.9, 096005 (2024)
[arXiv:2408.04589 [hep-ph]].




\bibitem{SPT-3G:2025bzu}
E.~Camphuis \textit{et al.} [SPT-3G],
arXiv:2506.20707 [astro-ph.CO].

\bibitem{rlimit}
P.~A.~R.~Ade \textit{et al.} [BICEP and Keck Collaborations],
Phys. Rev. Lett. \textbf{127}, no.15, 151301 (2021)
[arXiv:2110.00483 [astro-ph.CO]].



\bibitem{Tristram:2021tvh}
M.~Tristram, A.~J.~Banday, K.~M.~G\'orski, R.~Keskitalo, C.~R.~Lawrence, K.~J.~Andersen, R.~B.~Barreiro, J.~Borrill, L.~P.~L.~Colombo and H.~K.~Eriksen, \textit{et al.}
arXiv:2112.07961 [astro-ph.CO].


\bibitem{LiddleLeach} 
  A.~R.~Liddle and S.~M.~Leach,
  Phys.\ Rev.\ D {\bf 68}, 103503 (2003)
  [astro-ph/0305263];
  
\bibitem{Martin:2010kz}
J.~Martin and C.~Ringeval,
Phys. Rev. D \textbf{82}, 023511 (2010)
[arXiv:1004.5525 [astro-ph.CO]].

\bibitem{Ellis:2021kad}
J.~Ellis, M.~A.~G.~Garcia, D.~V.~Nanopoulos, K.~A.~Olive and S.~Verner,
Phys. Rev. D \textbf{105}, no.4, 043504 (2022)
[arXiv:2112.04466 [hep-ph]].







\bibitem{Pedro:2013qga}
F.~G.~Pedro, M.~Rummel and A.~Westphal,
[arXiv:1306.1237 [hep-th]].

\bibitem{Berg:2005yu}
M.~Berg, M.~Haack and B.~Kors,
Phys. Rev. Lett. \textbf{96}, 021601 (2006)
[arXiv:hep-th/0508171 [hep-th]].

\bibitem{vonGersdorff:2005bf}
G.~von Gersdorff and A.~Hebecker,
Phys. Lett. B \textbf{624}, 270-274 (2005)
[arXiv:hep-th/0507131 [hep-th]].

\bibitem{Ellis:1984bs}
J.~R.~Ellis, C.~Kounnas and D.~V.~Nanopoulos,
Phys. Lett. B \textbf{143}, 410-414 (1984)

\bibitem{Ellis:2015kqa}
J.~Ellis, M.~A.~G.~Garcia, D.~V.~Nanopoulos and K.~A.~Olive,
JCAP \textbf{10}, 003 (2015)
[arXiv:1503.08867 [hep-ph]].

\bibitem{Dudas:2017kfz}
E.~Dudas, T.~Gherghetta, Y.~Mambrini and K.~A.~Olive,
Phys. Rev. D \textbf{96}, no.11, 115032 (2017)
[arXiv:1710.07341 [hep-ph]].

\bibitem{Ema:2024sit}
Y.~Ema, M.~A.~G.~Garcia, W.~Ke, K.~A.~Olive and S.~Verner,
Universe \textbf{10}, no.6, 239 (2024)
[arXiv:2404.14545 [hep-ph]].

\end{thebibliography}

\end{document}